\documentclass[journal]{IEEEtran}
\usepackage{cite}
\usepackage{graphicx}
\usepackage{amssymb,amsmath,bm}
\usepackage[colorlinks,linkcolor=black,anchorcolor=black,citecolor=black,urlcolor=black]{hyperref}
\usepackage{hyperref}
\usepackage{multirow}
\usepackage{amsfonts}
\usepackage{url}
\usepackage{multirow}
\usepackage{siunitx}
\usepackage{color}
\usepackage{tabularx}
\usepackage{color}
\usepackage{makecell}
\ifCLASSINFOpdf
\else
\fi
\begin{document}
%
\title{APNet: An All-Frame-Level Neural Vocoder Incorporating Direct Prediction of Amplitude and Phase Spectra}
%
%
%

\author{Yang~Ai,~\IEEEmembership{Member,~IEEE},~Zhen-Hua~Ling,~\IEEEmembership{Senior Member,~IEEE}
\thanks{This work was partially supported by the National Natural Science Foundation of China under Grant 61871358 and the Fundamental Research Funds for the Central Universities.}
\thanks{Y. Ai and Z.-H. Ling are with the National Engineering Research Center of Speech and Language Information Processing, University of Science and Technology of China, Hefei, 230027, China (e-mail: yangai@ustc.edu.cn, zhling@ustc.edu.cn).}
}
%
%

\markboth{}%
{Shell \MakeLowercase{\textit{et al.}}: Bare Demo of IEEEtran.cls for Journals}
%



\maketitle

\begin{abstract}
This paper presents a novel neural vocoder named APNet which reconstructs speech waveforms from acoustic features by predicting amplitude and phase spectra directly.
The APNet vocoder is composed of an amplitude spectrum predictor (ASP) and a phase spectrum predictor (PSP).
The ASP is a residual convolution network which predicts frame-level log amplitude spectra from acoustic features.
The PSP also adopts a residual convolution network using acoustic features as input, then passes the output of this network through two parallel linear convolution layers respectively, and finally integrates into a phase calculation formula to estimate frame-level phase spectra.
Finally, the outputs of ASP and PSP are combined to reconstruct speech waveforms by inverse short-time Fourier transform (ISTFT).
All operations of the ASP and PSP are performed at the frame level.
We train the ASP and PSP jointly and define multilevel loss functions based on amplitude mean square error, phase anti-wrapping error, short-time spectral inconsistency error and time domain reconstruction error.
Experimental results show that our proposed APNet vocoder achieves an approximately 8x faster inference speed than HiFi-GAN v1 on a CPU due to the all-frame-level operations, while its synthesized speech quality is comparable to HiFi-GAN v1.
The synthesized speech quality of the APNet vocoder is also better than that of several equally efficient models.
Ablation experiments also confirm that the proposed parallel phase estimation architecture is essential to phase modeling and the proposed loss functions are helpful for improving the synthesized speech quality.

\end{abstract}


%
\IEEEpeerreviewmaketitle

\section{Introduction}

\IEEEPARstart{S}
{peech} synthesis, also known as text-to-speech (TTS), converts any input text into natural and fluent speech and solves the problem of how to make machines speak like humans.
A speech synthesis system with high naturalness, intelligibility and expressiveness is a goal pursued by speech synthesis researchers.
In recent years, statistical parametric speech synthesis (SPSS) \cite{zen2009statistical} has become the mainstream speech synthesis method.
It has the advantages of small system size, high robustness and flexibility.
The acoustic model and vocoder are two key modules in the SPSS framework.
The acoustic model describes the mapping relationship from text to acoustic features and currently incorporates many advanced structures, such as Tacotron \cite{wang2017tacotron}, Tacotron2 \cite{shen2018natural}, and Transformer \cite{li2019neural}, which outperform the conventional hidden Markov model (HMM) \cite{tokuda2013speech}, deep neural network (DNN) \cite{ze2013statistical} and recurrent neural network (RNN) \cite{fan2014tts} based methods.
Vocoders \cite{dudley1939vocoder}, which reconstruct speech waveforms from acoustic features (e.g., mel spectrograms), also play an important role in SPSS.
Their performance significantly affects the quality of synthetic speech.
Some conventional vocoders (e.g., STRAIGHT \cite{kawahara1999restructuring} and WORLD \cite{morise2016world}) have some deficiencies, such as the loss of spectral details and phase information.
Thanks to the development of deep learning and neural networks, \emph{neural vocoders} have emerged and gradually replaced the conventional ones.
Neural vocoders can also be applied to other tasks, such as voice conversion \cite{liu2018wavenet}, bandwidth extension \cite{liu2022neural} and speech enhancement \cite{andreev2022hifi++}.

The synthesized speech quality and inference efficiency are the two main considerations for evaluating neural vocoders.
Initially, autoregressive neural waveform generation models represented by WaveNet \cite{oord2016wavenet}, SampleRNN \cite{mehri2016samplernn} and WaveRNN \cite{kalchbrenner2018efficient} were proposed and successfully applied in the construction of neural vocoders \cite{tamamori2017speaker,hayashi2017investigation,adiga2018use,ai2018samplernn,ai2019dnn,lorenzo2018robust}.
Although autoregressive neural vocoders achieved breakthroughs in synthesized speech quality compared to conventional ones,
their inference efficiency was unsatisfactory due to the autoregressive inference mode of raw waveforms with a high sampling rate.
Subsequently, knowledge-distilling-based models (e.g., Parallel WaveNet \cite{oord2017parallel} and ClariNet \cite{ping2018clarinet}), flow-based models (e.g., WaveGlow \cite{prenger2018waveglow} and WaveFlow \cite{ping2020waveflow}), and glottis-based models (e.g., GlotNet \cite{juvela2019glotnet} and LPCNet \cite{valin2019lpcnet}) were proposed.
Although the inference efficiency of these models has improved significantly, the overall computational complexity was still high, limiting their use in resource-constrained environments such as embedded devices.
Recently, waveform generation models without autoregressive and flow-like structures have received more attention.
The neural source-filter (NSF) model \cite{wang2019neural} combined speech production mechanisms with neural networks and achieved the prediction of speech waveforms from explicit F0 and mel spectrograms.
Generative adversarial network (GAN) \cite{goodfellow2014generative} based models, such as WaveGAN \cite{donahue2018adversarial}, GAN-TTS \cite{binkowski2019high}, MelGAN \cite{kumar2019melgan}, Parallel WaveGAN \cite{yamamoto2020parallel} and HiFi-GAN \cite{kong2020hifi}, leveraged GANs to ensure the synthesized speech quality and improved the efficiency by parallelizable training and inference through noncausal convolutions.
For example, HiFi-GAN \cite{kong2020hifi} cascades multiple upsampling layers and residual convolution networks to gradually upsample the input mel spectrogram to the sampling rate of the final waveform while performing convolution operations.
It also adopts adversarial training to ensure high-fidelity waveform generation.
However, these GAN-based models always directly predict the final waveforms, so the problem of slow inference still exists.

A speech waveform can be interconverted with its amplitude and phase spectra.
Predicting the amplitude and phase spectra instead of the waveform from the input acoustic features is a new approach to build neural vocoders.
To improve the inference efficiency of HiFi-GAN, ISTFTNET \cite{kaneko2022istftnet} proposes to output the amplitude and phase spectra close to the waveform sampling rate in the middle layer of HiFi-GAN and then synthesize the final waveform through inverse short-time Fourier transform (ISTFT).
However, the results reported in ISTFTNET \cite{kaneko2022istftnet} show that predicting amplitude and phase spectra at low sampling rates is difficult.
In our previous work, we proposed a neural vocoder named HiNet \cite{ai2020neural} which reconstructs speech waveforms from acoustic features by predicting amplitude and phase spectra hierarchically.
The predicted spectra are at the same sampling rate as the acoustic features.
Unfortunately, limited by the issue of phase wrapping and the difficulty of phase modeling, the HiNet vocoder still needs an NSF-like model to generate an intermediate waveform for further phase extraction through short-time Fourier transform (STFT).
To our knowledge, synthesizing a waveform by predicting its raw amplitude and phase spectra directly has not yet been thoroughly investigated.

Therefore, this paper proposes a novel neural vocoder named APNet, which incorporates the direct prediction of amplitude and phase spectra.
Similar to the HiNet vocoder, the proposed APNet vocoder consists of an amplitude spectrum predictor (ASP) and a phase spectrum predictor (PSP).
The ASP and PSP predict the raw amplitude and phase spectra at the same sampling rate as the acoustic features, respectively, and then the final waveform is synthesized through ISTFT.
Compared to the HiNet vocoder, the APNet vocoder 1) jointly trains the ASP and PSP and 2) predicts the phase spectra directly.
The ASP is a residual convolution network which predicts frame-level log amplitude spectra from acoustic features.
The PSP also adopts a residual convolution network using acoustic features as input.
Then, the output of the network is passed through two parallel linear convolution layers respectively to obtain two outputs, and then the frame-level phase spectra are estimated by a phase calculation formula.
Therefore, all operations in APNet are performed at the frame level, expecting to significantly improve the inference efficiency.
We also propose multilevel loss functions defined on predicted log amplitude spectra, phase spectra, reconstructed STFT spectra and final waveforms to ensure the fidelity of the amplitude spectra, the precision of the phase spectra, the consistency of the STFT spectra and the quality of the final waveforms.
Experimental results show that the synthesized speech quality of our proposed APNet vocoder is comparable to that of HiFi-GAN v1 \cite{kong2020hifi}, but its inference efficiency is significantly improved due to the all-frame-level operations, reaching 14x real-time on a CPU, which is 8x faster than HiFi-GAN v1.
Although the inference efficiency on a CPU of multi-band MelGAN \cite{yang2021multi} and ISTFTNET \cite{kaneko2022istftnet} is similar to our model, their synthesized speech quality is clearly inferior to our model.
The APNet vocoder outperforms the HiNet vocoder in terms of both synthesized speech quality and inference efficiency.
We also conduct several ablation experiments to explore the roles of specific structures and loss functions.
Interestingly, the proposed parallel phase estimation architecture is the key to successful phase modeling.
The proposed loss functions also help to improve the synthesized speech quality.

There are two main contributions of the APNet vocoder.
First, the APNet achieves all-frame-level speech waveform generation, which greatly improves the inference efficiency by predicting the frame-level amplitude and phase spectra.
Second, the proposed parallel phase estimation architecture in APNet successfully achieves phase modeling using neural networks, which provides a new idea for the tricky phase estimation task.

This paper is organized as follows: In Section \ref{sec: Related works}, we briefly review the HiFi-GAN \cite{kong2020hifi}, ISTFTNET \cite{kaneko2022istftnet} and our previously proposed HiNet vocoder \cite{ai2020neural}, respectively.
In Section \ref{sec: Proposed Methods}, we provide details on our proposed APNet vocoder. In Section \ref{sec: Experiments}, we present our experimental results. Finally, we give conclusions in Section \ref{sec: Conclusion}.

\section{Related Works}
\label{sec: Related works}

\begin{figure*}
    \centering
    \includegraphics[height=7cm]{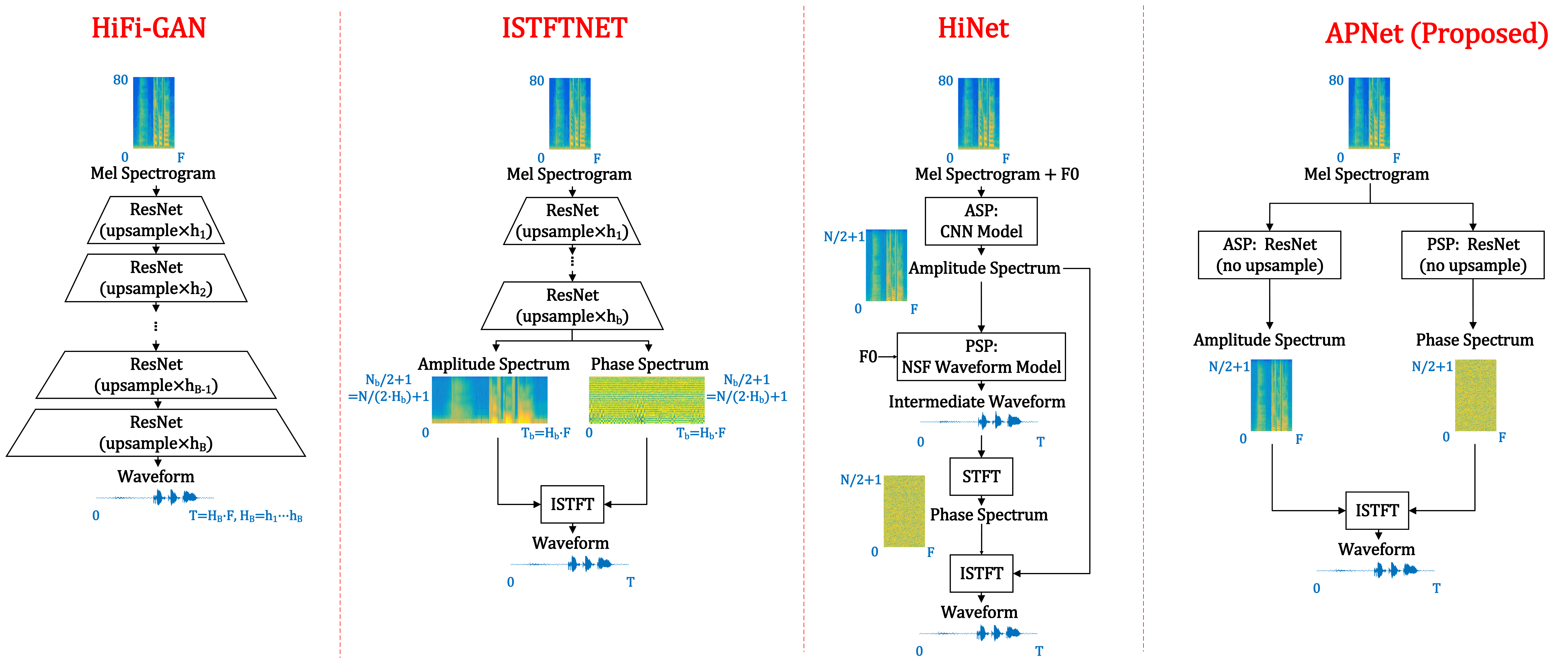}
    \caption{Model structure comparison among the HiFi-GAN, ISTFTNET, HiNet and APNet vocoders. Here, \emph{ASP} and \emph{PSP} represent the amplitude spectrum predictor and phase spectrum predictor, respectively. \emph{ResNet}, \emph{CNN} and \emph{NSF} represent the residual convolution network, convolutional neural network and neural source-filter, respectively. \emph{STFT} and \emph{ISTFT} represent the short-time Fourier transform and inverse short-time Fourier transform, respectively.}
    \label{fig: Comparsion}
\end{figure*}

In this section, we briefly introduce the HiFi-GAN \cite{kong2020hifi}, ISTFTNET \cite{kaneko2022istftnet} and HiNet vocoders \cite{ai2020neural}, respectively.
These vocoders are used for comparison with our proposed APNet vocoder in the experiments.
Figure \ref{fig: Comparsion} shows a model structure comparison among the four vocoders.

\subsection{HiFi-GAN}
\label{subsec2A: HiFi-GAN}

As shown in Figure \ref{fig: Comparsion}, HiFi-GAN \cite{kong2020hifi} cascades $B$ residual convolution networks, converting 80-dimensional frame-level mel spectrogram $\bm{M}\in \mathbb{R}^{F\times 80}$ with length $F$ to waveform $\hat{\bm{x}}\in \mathbb{R}^T$ with length $T$.
$\frac{T}{F}$ is also equal to the frame shift when extracting the mel spectrogram from a waveform.
The residual convolution network includes dilated convolution, residual connections, skip connections, etc.
There is an upsampling layer in front of each convolutional network, gradually raising the length $F$ of $\bm{M}$ to the length $T$ of $\hat{\bm{x}}$.
Suppose the upsampling ratio of the $b$-th residual convolution network is $h_b$, where $b=1,2,\dots,B$, then we can obtain $T=H_B\cdot F$, where $H_b=h_1\cdot h_2\cdots h_b, b=1,\dots,B$.
Assume that the sampling rate of the waveform is $f_s$.
Obviously, HiFi-GAN operates at multiple sampling rates, which are $F\cdot\frac{f_s}{T},H_1F\cdot\frac{f_s}{T},\dots,H_{B-1}F\cdot\frac{f_s}{T},f_s$.
HiFi-GAN adopts adversarial training with multi-period discriminator (MPD) and multi-scale discriminator (MSD) to ensure high-fidelity waveform generation.
The final loss of HiFi-GAN is a combination of GAN loss, feature matching loss and mel spectrogram loss.

\subsection{ISTFTNET}
\label{subsec2B: ISTFTNET}

To improve the inference efficiency, ISTFTNET \cite{kaneko2022istftnet} incorporates ISTFT into HiFi-GAN.
As shown in Figure \ref{fig: Comparsion}, ISTFTNET outputs the amplitude spectrum $\hat{\bm{A}}\in \mathbb{R}^{T_b\times (N_b/2+1)}$ and phase spectrum $\hat{\bm{P}}\in \mathbb{R}^{T_b\times (N_b/2+1)}$ with sampling rate $H_{b}F\cdot\frac{f_s}{T}$ after the intermediate $b$-th ($1<b<B$) residual convolution network of HiFi-GAN and then synthesizes the final waveform through ISTFT.
Although the predicted amplitude and phase spectra are of length $T_b=H_b\cdot F >F $, their FFT point number is $N_b=\frac{N}{H_b}<N$, where $N$ is the FFT point number when extracting the mel spectrogram from a waveform.
That is, ISTFTNET predicts amplitude and phase spectra with high time resolution and low frequency resolution.
Therefore, ISTFTNET operates at multiple sampling rates, which are $F\cdot\frac{f_s}{T}, H_1F\cdot\frac{f_s}{T},\dots,H_{b}F\cdot\frac{f_s}{T}$, and hence achieves higher inference efficiency than HiFi-GAN.
However, the results reported in ISTFTNET \cite{kaneko2022istftnet} show that predicting amplitude and phase spectra at low sampling rates is tricky, i.e., $T_b$ cannot be too small than $T$.

\subsection{HiNet}
\label{subsec2C: HiNet}

In our previous work, we proposed the HiNet vocoder \cite{ai2020neural} and its variant \cite{ai2020knowledge}.
We have also successfully applied the HiNet vocoder in the reverberation modeling task \cite{ai2020reverberation} and denoising and dereveberation task \cite{ai2021denoising,ai2022denoising}, respectively.
As shown in Figure \ref{fig: Comparsion}, the HiNet vocoder uses an ASP and a PSP to predict the frame-level log amplitude spectrum and phase spectrum of a waveform, respectively.
During the inference process, the ASP predicts the log amplitude spectrum $\log\hat{\bm{A}}\in \mathbb{R}^{F\times (N/2+1)}$ from the input mel spectrogram and F0, and the PSP predicts the phase spectrum $\hat{\bm{P}}\in \mathbb{R}^{F\times (N/2+1)}$ from the input F0 and predicted log amplitude spectrum by ASP.
The length and FFT point number of $\hat{\bm{A}}$ and $\hat{\bm{P}}$ are $F$ and $N$ respectively, which are the same as those of the input mel spectrogram.
It then reconstructs the waveform from the predicted amplitude and phase spectra through ISTFT.
The ASP is a frame-level convolutional neural network (CNN) containing multiple convolutional layers.
However, limited by the difficulty of phase modeling, the PSP is constructed by concatenating a lightweight NSF-like neural waveform model with a phase spectrum extractor.
The NSF-like model predicts an intermediate waveform $\hat{\bm{x}}_I\in \mathbb{R}^T$ and then the extractor extracts the phase spectrum from $\hat{\bm{x}}_I$ through STFT.
HiNet operates at two sampling rates which are $F\cdot\frac{f_s}{T}$ and $f_s$, and still fails to achieve all-frame-level waveform generation at the unique raw sampling rate $F\cdot\frac{f_s}{T}$.

\section{Proposed Methods}
\label{sec: Proposed Methods}

\begin{figure*}
    \centering
    \includegraphics[height=7cm]{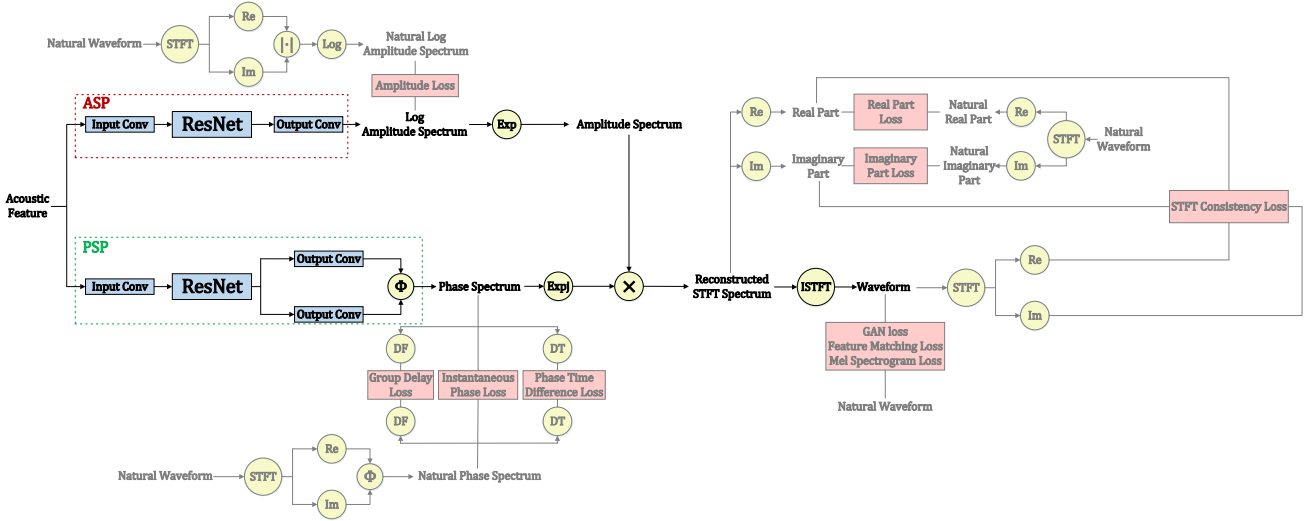}
    \caption{Details of the APNet vocoder. Here, \emph{ASP} and \emph{PSP} represent the amplitude spectrum predictor and phase spectrum predictor, respectively. Blue boxes, red boxes and yellow circles represent trainable modules, loss functions and basic operations, respectively. \emph{ResBlock} and \emph{Conv} represent the residual convolution network and linear convolution layer, respectively. \emph{STFT}, \emph{ISTFT}, \emph{Re}, \emph{Im}, \emph{$|\cdot|$}, \emph{$\Phi$}, \emph{DF}, \emph{DT}, \emph{Exp}, \emph{Log}, \emph{Expj} and \emph{$\times$} represent the short-time Fourier transform, inverse short-time Fourier transform, real part calculation, imaginary part calculation, amplitude calculation, phase calculation, differential along frequency axis, differential along time axis, natural exponential function, logarithm, complex exponential function $e^{j\cdot}$ and element-wise multiplication, respectively. \emph{GAN} denotes generative adversarial network. Gray parts do not appear during inference.
    }
    \label{fig: APNet}
\end{figure*}

As shown in Figure \ref{fig: Comparsion}, the proposed APNet vocoder directly predicts the speech amplitude spectrum $\hat{\bm{A}}\in \mathbb{R}^{F\times (N/2+1)}$ and phase spectrum $\hat{\bm{P}}\in \mathbb{R}^{F\times (N/2+1)}$ at unique raw sampling rate $F\cdot\frac{f_s}{T}$ from input acoustic feature (e.g., mel spectrogram $\bm{M}\in \mathbb{R}^{F\times 80}$).
Then, the amplitude and phase spectra are reconstructed to the STFT spectrum $\hat{\bm{S}}\in \mathbb{C}^{F\times (N/2+1)}$, and finally, a waveform $\hat{\bm{x}}\in \mathbb{R}^T$ is recovered through ISTFT.
The whole process can be described by the following mathematical expression:
\begin{align}
\label{equ: whole process1}
\log\hat{\bm{A}}=ASP(\bm{M}),\\
\label{equ: whole process2}
\hat{\bm{P}}=PSP(\bm{M}),\\
\label{equ: whole process3}
\hat{\bm{S}}=\hat{\bm{A}}\cdot e^{j\hat{\bm{P}}},\\
\label{equ: whole process4}
\hat{\bm{x}}=ISTFT(\hat{\bm{S}}).
\end{align}
Therefore, the APNet vocoder is an all-frame-level model and only operates at sampling rate $F\cdot\frac{f_s}{T}$.
Detailed structure and loss functions are shown in Figure \ref{fig: APNet}.
In Figure \ref{fig: APNet}, only the blue boxes are trainable, and the gray part disappears during the inference process.

\begin{figure*}
    \centering
    \includegraphics[height=2.5cm]{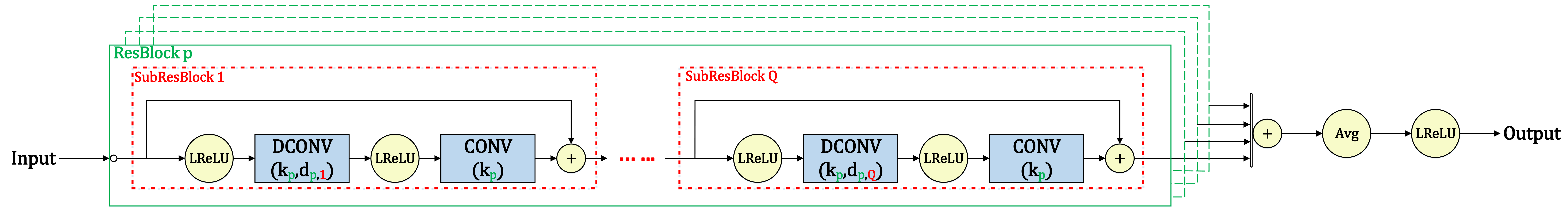}
    \caption{Details of the residual convolution network used in the ASP and PSP. Here, \emph{CONV}, \emph{DCONV}, \emph{ResBlock} and \emph{SubResBlock} represent the linear convolution layer, linear dilated convolution layer, residual convolution block and residual convolution subblock, respectively. \emph{LReLU}, \emph{Avg} and \emph{$+$} represent leaky rectified linear unit, average and addition, respectively. $k_*$ and $d_{*,*}$ denotes kernel size and dilation factor, respectively.
    }
    \label{fig: ResNet}
\end{figure*}

\subsection{Amplitude Spectrum Predictor}
\label{subsec3A: ASP}

As shown in Figure \ref{fig: APNet}, the ASP predicts the log amplitude spectrum $\log\hat{\bm{A}}$ from input acoustic feature (i.e., Equation \ref{equ: whole process1}).
It consists of an input linear convolution layer, a residual convolution network and an output linear convolution layer.
The details of the residual convolution network are shown in Figure \ref{fig: ResNet}.
The residual convolution network consists of $P$ parallel residual convolution blocks (i.e., \emph{ResBlock} in Figure \ref{fig: ResNet}), all of which have the same input.
Then, the outputs of these $P$ blocks are summed (i.e., skip connections), averaged (i.e., divided by $P$), and finally activated by a leaky rectified linear unit (LReLU) \cite{maas2013rectifier}.
The $p$-th block ($p=1,\dots,P$) is formed by concatenating $Q$ residual convolution subblocks (i.e., \emph{SubResBlock} in Figure \ref{fig: ResNet}).
In the $q$-th subblock ($q=1,\dots,Q$) of the $p$-th block ($p=1,\dots,P$), the input is first activated by LReLU, then passes through a linear dilated convolution layer (kernel size $= k_p$ and dilation factor $= d_{p,q}$), then is activated by LReLU again, passes through a normal linear convolution layer (kernel size $= k_p$), and finally superimposes with the input (i.e., residual connections) to obtain the output.

\subsection{Phase Spectrum Predictor}
\label{subsec3B: PSP}

As shown in Figure \ref{fig: APNet}, the PSP predicts the phase spectrum $\hat{\bm{P}}$ from input acoustic feature (i.e., Equation \ref{equ: whole process2}).
It consists of an input linear convolution layer, a residual convolution network, two parallel output linear convolution layers and a phase calculation formula $\bm{\Phi}$.
The residual convolution network is exactly the same as the one used in ASP.
Assuming that two parallel output layers output $\bm{\widetilde R}$ and $\bm{\widetilde I}$ respectively, then
\begin{align}
\label{equ: Phase_calculate_matrix}
\hat{\bm{P}}=\bm{\Phi}(\bm{\widetilde R},\bm{\widetilde I}).
\end{align}
Note that Equation \ref{equ: Phase_calculate_matrix} is calculated element-wise, i.e., $\hat{P}_{f,n}=\bm{\Phi}(\widetilde R_{f,n},\widetilde I_{f,n}$), $f=1,\dots,F, n=1,\dots,\frac{N}{2}+1$, where $\hat{P}_{f,n}$, $\widetilde R_{f,n}$ and $\widetilde I_{f,n}$ are the elements in $\hat{\bm{P}}$, $\bm{\widetilde R}$ and $\bm{\widetilde I}$, respectively.
The range of values for the phase is $-\pi<\hat{P}_{f,n}\leq\pi$.

\begin{figure}
    \centering
    \includegraphics[height=6cm]{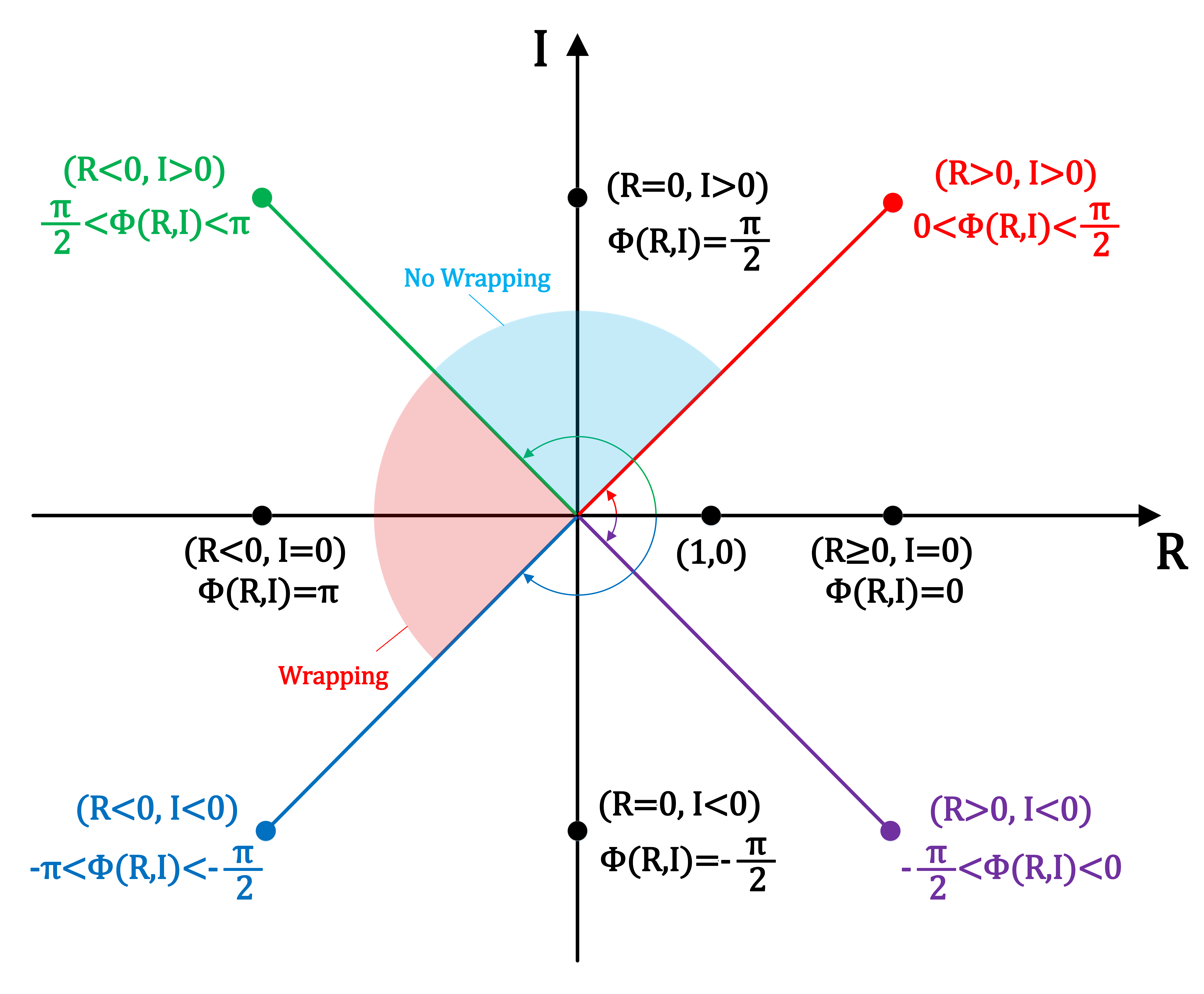}
    \caption{A quadrant diagram used to explain the phase calculation formula $\bm{\Phi}(R,I)$ and the phase wrapping issue.
    }
    \label{fig: Phase_calculate}
\end{figure}

We derive the mathematical expression for phase calculation formula $\bm{\Phi}$.
First, we define two real numbers $R\in\mathbb{R}$ and $I\in\mathbb{R}$, and define a new symbolic function
\begin{align}
\label{equ: Symbolic_function}
Sgn^*(x)=\left\{\begin{array}{rl}1,& x\ge 0\\ -1,&x<0\end{array}\right..
\end{align}
The phase calculation formula $\bm{\Phi}$ is a function with two variables.
Since $\arctan(-\infty)=-\frac{\pi}{2}$ and $\arctan(+\infty)=\frac{\pi}{2}$, $\bm{\Phi}(R,I)$ can be written by arctangent and symbolic functions as follows
\begin{align}
\label{equ: Phase calculation}
\bm{\Phi}(R,I)=\arctan\left(\dfrac{I}{R}\right)-\dfrac{\pi}{2}\cdot Sgn^*(I)\cdot\left[Sgn^*(R)-1\right],
\end{align}
and sets $\bm{\Phi}(0,0)=0$.
For $\forall R\in\mathbb{R}$ and $I\in\mathbb{R}$, $-\pi<\bm{\Phi}(R,I)\leq\pi$ (i.e., the resulting phase value is restricted within the principal value interval $(-\pi,\pi]$).
To visually explain Equation \ref{equ: Phase calculation}, we draw a quadrant diagram shown in Figure \ref{fig: Phase_calculate}.
The horizontal and vertical axes represent $R$ and $I$, respectively.
Figure \ref{fig: Phase_calculate} reveals the value range of $\bm{\Phi}(R,I)$ when $R$ and $I$ take different values.
When the vector $(R,I)$ is located at the origin or on the real horizontal axis, $\bm{\Phi}(R,I)=0$;
when the vector $(R,I)$ is in the first and second quadrants (including the positive vertical axis and negative horizontal axis), $\bm{\Phi}(R,I)$ is the angle through which the vector $(1,0)$ rotates counterclockwise to the vector $(R,I)$;
when the vector $(R,I)$ is in the third and fourth quadrants (including the negative vertical axis), $\bm{\Phi}(R,I)$ is the negative angle through which the vector $(1,0)$ rotates clockwise to the vector $(R,I)$.

We name these two parallel output layers and the phase calculation formula $\bm{\Phi}$ as \emph{parallel phase estimation architecture}.
The proposed architecture is the key to successful phase modeling, which is confirmed in the following experiments.

\subsection{Training and Inference Process}
\label{subsec3C: Training and Inference Process}

At the training stage, we jointly train the ASP and PSP and define multilevel loss functions on predicted log amplitude spectra, phase spectra, reconstructed STFT spectra and final waveforms as shown in the red boxes in Figure \ref{fig: APNet}.
These loss functions are expected to ensure the fidelity of the amplitude spectra, the precision of the phase spectra, the consistency of the STFT spectra and the quality of the final waveforms.
These loss functions are described below.

\subsubsection{Losses defined on log amplitude spectra}
\label{subsubsec3C1: Losses defined on log amplitude spectra}

The losses defined on log amplitude spectra are used to close the gap between the predicted log amplitude spectra and the natural ones, facilitating the generation of high-quality and high-fidelity log amplitude spectra, including:
\begin{itemize}
\item \textbf{Amplitude Loss}: The amplitude loss is the mean square error (MSE) between the predicted log amplitude spectrum $\log\hat{\bm{A}}$ and the natural one $\log\bm{A}$, i.e.,
    \begin{align}
    \label{equ: Amplitude Loss}
    \mathcal L_{A}=\mathbb{E}_{\left(\log\hat{\bm{A}},\log\bm{A}\right)}\overline{\left(\log\hat{\bm{A}}-\log \bm{A} \right)^2},
    \end{align}
    where $\overline{\bm{X}}$ means averaging all elements in the matrix $\bm{X}$.
    Here, to extract the natural log amplitude spectrum, we first perform STFT on the natural speech waveform $\bm{x}$ to obtain the STFT spectrum $\bm{S}$, i.e.,
    \begin{align}
    \label{equ: STFT}
    \bm{S}=STFT\left(\bm{x};L,\frac{T}{F},N\right),
    \end{align}
    where $L$, $\frac{T}{F}$ and $N$ are the frame length, frame shift and FFT point number, respectively.
    Then, the natural log amplitude spectrum is calculated by the following amplitude calculation formula
    \begin{align}
    \label{equ: natural amplitude calculation formula}
    \log\bm{A}=\log\sqrt{Re^2(\bm{S})+Im^2(\bm{S})},
    \end{align}
    where $Re$ and $Im$ are the real part calculation and imaginary part calculation, respectively.

\end{itemize}

\subsubsection{Losses defined on phase spectra}
\label{subsubsec3C2: Losses defined on phase spectra}

The losses defined on phase spectra are well designed to overcome the phase wrapping issue and used to close the gap between the predicted phase spectra and the natural ones, expecting to improve the precision and continuity of phase spectra, including:

\begin{itemize}
\item \textbf{Instantaneous Phase Loss}: The instantaneous phase loss is the negative cosine loss between the predicted instantaneous phase spectrum $\hat{\bm{P}}$ and the natural one $\bm{P}$, i.e.,
    \begin{align}
    \label{equ: Phase Loss}
    \mathcal L_{IP}=-\mathbb{E}_{\left(\hat{\bm{P}},\bm{P}\right)} \overline{\cos\left(\hat{\bm{P}}-\bm{P} \right)}.
    \end{align}
    The natural instantaneous phase spectrum is calculated by Equation \ref{equ: Phase calculation} as follows:
    \begin{align}
    \label{equ: natural phase calculation formula}
    \bm{P}=\bm{\Phi}\left(Re(\bm{S}),Im(\bm{S})\right).
    \end{align}

    \item \textbf{Group Delay Loss}: Group delay is defined as the negative derivative of phase by frequency, reflecting phase variations in neighboring frequency bins.
    For the discrete phase spectrum, its group delay is the difference along the frequency axis.
    Define the group delay matrix as follows:
    \begin{align}
    \label{equ: group delay matrix}
    \bm{W}=\left[\bm{w}_1,\dots,\bm{w}_n,\dots,\bm{w}_{\frac{N}{2}+1}\right],\\
    \bm{w}_n=\left[\mathop{0}_{\text{1st}},\dots,\mathop{0}_{},\mathop{1}_{n\text{-th}},\mathop{-1}_{},\mathop{0}_{},\dots,
    \mathop{0}_{\left(\frac{N}{2}+1\right)\text{-th}} \right]^\top.
    \end{align}
    Therefore, we can calculate the group delay of $\hat{\bm{P}}$ and $\bm{P}$ by
    \begin{align}
    \label{equ: group delay calculate}
    \Delta_{DF}\hat{\bm{P}}=\hat{\bm{P}}\bm{W},\\
    \Delta_{DF}\bm{P}=\bm{P}\bm{W}.
    \end{align}
    The group delay loss is the negative cosine loss between the predicted group delay $\Delta_{DF}\hat{\bm{P}}$ and the natural one $\Delta_{DF}\bm{P}$, i.e.,
    \begin{align}
    \label{equ: Group Delay Loss}
    \mathcal L_{GD}=-\mathbb{E}_{\left(\Delta_{DF}\hat{\bm{P}},\Delta_{DF}\bm{P}\right)} \overline{\cos\left(\Delta_{DF}\hat{\bm{P}}-\Delta_{DF}\bm{P} \right)}.
    \end{align}
    The group delay loss narrows the difference in continuity of natural and predicted phases along the frequency axis.

    \item \textbf{Phase Time Difference Loss}: Similarly, we define the phase time difference matrix
    \begin{align}
    \label{equ: PTD matrix}
    \bm{V}=\left[\bm{v}_1,\dots,\bm{v}_f,\dots,\bm{v}_F\right]^\top,\\
    \bm{v}_f=\left[\mathop{0}_{\text{1st}},\dots,\mathop{0}_{},\mathop{1}_{f\text{-th}},\mathop{-1}_{},\mathop{0}_{},\dots,
    \mathop{0}_{F\text{-th}} \right]^\top,
    \end{align}
    and calculate the time difference of $\hat{\bm{P}}$ and $\bm{P}$ as follows:
    \begin{align}
    \label{equ: PTD calculate}
    \Delta_{DT}\hat{\bm{P}}=\bm{V}\hat{\bm{P}},\\
    \Delta_{DT}\bm{P}=\bm{V}\bm{P}.
    \end{align}
    The phase time difference loss is the negative cosine loss between the predicted phase time difference $\Delta_{DT}\hat{\bm{P}}$ and the natural one $\Delta_{DT}\bm{P}$, i.e.,
    \begin{align}
    \label{equ: PTD Loss}
    \mathcal L_{PTD}=-\mathbb{E}_{\left(\Delta_{DT}\hat{\bm{P}},\Delta_{DT}\bm{P}\right)} \overline{\cos\left(\Delta_{DT}\hat{\bm{P}}-\Delta_{DT}\bm{P} \right)}.
    \end{align}
    The phase time difference loss narrows the difference in continuity of natural and predicted phases along the time axis.
\end{itemize}

\begin{figure}
    \centering
    \includegraphics[height=3cm]{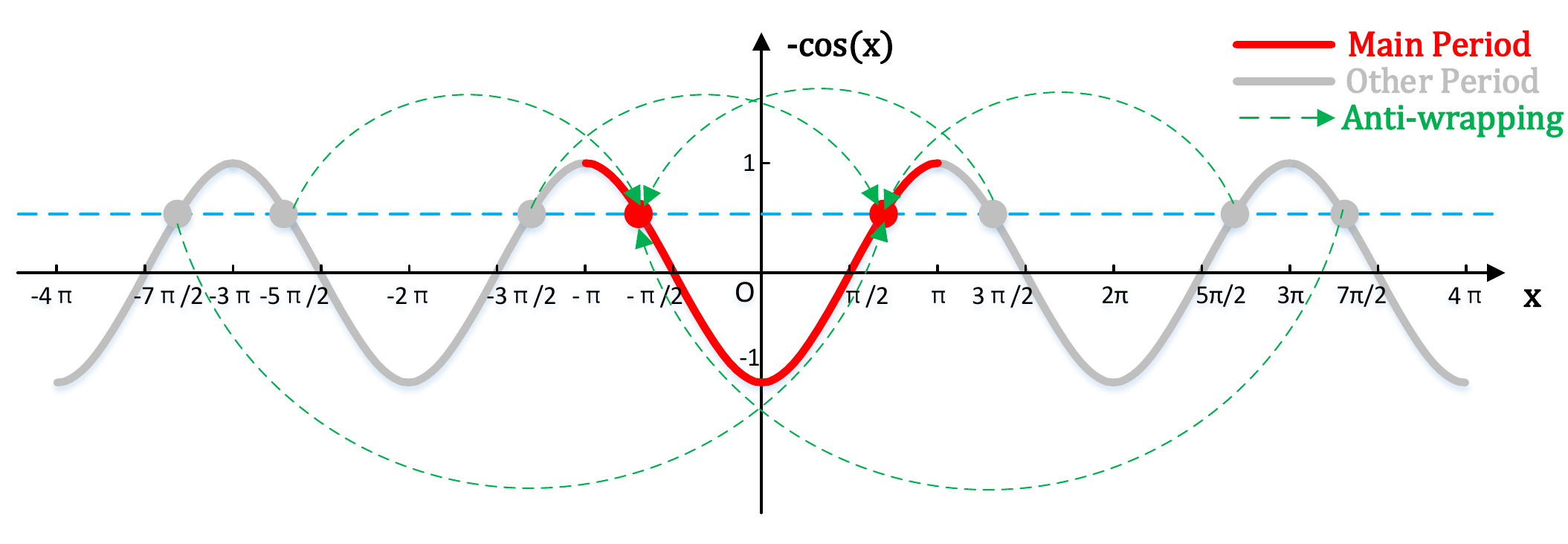}
    \caption{An interpretation diagram of the negative cosine function used as phase-related loss evaluation.
    }
    \label{fig: Negative_Cos}
\end{figure}

The total loss defined on phase spectra is
\begin{align}
\label{equ: phase loss_total}
\mathcal L_{P}=\mathcal L_{IP}+\mathcal L_{GD}+\mathcal L_{PDT}.
\end{align}

The negative cosine function used in Equation \ref{equ: Phase Loss}, \ref{equ: Group Delay Loss} and \ref{equ: PTD Loss} is suitable for the definition of phase-related losses because of its three properties\footnote{Theoretically, other functions that fit these three properties can also be used to evaluate phase-related losses.}, including:

\noindent \textbf{a) The parity}. The negative cosine function is an even function that only cares about the absolute value of the phase difference and ignores its sign.
This is reasonable since we expect the predicted phase to approximate the natural one regardless of the direction from which it is approached.

\noindent \textbf{b) The periodicity}.
The negative cosine function has a period of $2\pi$, which can potentially avoid the issue caused by phase wrapping.
As shown in Figure \ref{fig: Phase_calculate}, the absolute values of the phase difference represented by the blue and red shaded areas are equal.
However, in the actual calculation, the phase difference of the red shaded area (denoted by $P_{diff}^{wrapping}$) has a wrapping issue (i.e., $P_{diff}^{wrapping}\notin (-\pi,\pi]$) and its relationship with the phase difference of the blue shaded area (denoted by $P_{diff}^{no-wrapping}$) is
\begin{align}
\label{equ: phase wrapping}
|P_{diff}^{no-wrapping}|=2\pi-|P_{diff}^{wrapping}|.
\end{align}
Then we can get
\begin{align}
\label{equ: phase wrapping cos}
-\cos\left(P_{diff}^{no-wrapping}\right)=-\cos\left(P_{diff}^{wrapping}\right),
\end{align}
which indicates that the negative cosine function is an anti-wrapping function which can avoid the error caused by phase wrapping.
Therefore, as shown in Figure \ref{fig: Negative_Cos}, when calculating $\mathcal L_{IP}$, $\mathcal L_{GD}$ and $\mathcal L_{PTD}$, the negative cosine function potentially moves values outside $(-\pi,\pi]$ (i.e., gray dots) to the corresponding values within $(-\pi,\pi]$ (i.e., red dots), which performs anti-wrapping operation.

\noindent \textbf{c) The monotonicity}.
The negative cosine function is monotonically increasing in interval $[0,\pi]$, ensuring that the smaller the absolute value of the phase difference, the smaller the loss.

\subsubsection{Losses defined on reconstructed STFT spectra}
\label{subsubsec3C3: Losses defined on reconstructed STFT spectra}

\begin{figure}
    \centering
    \includegraphics[height=4cm]{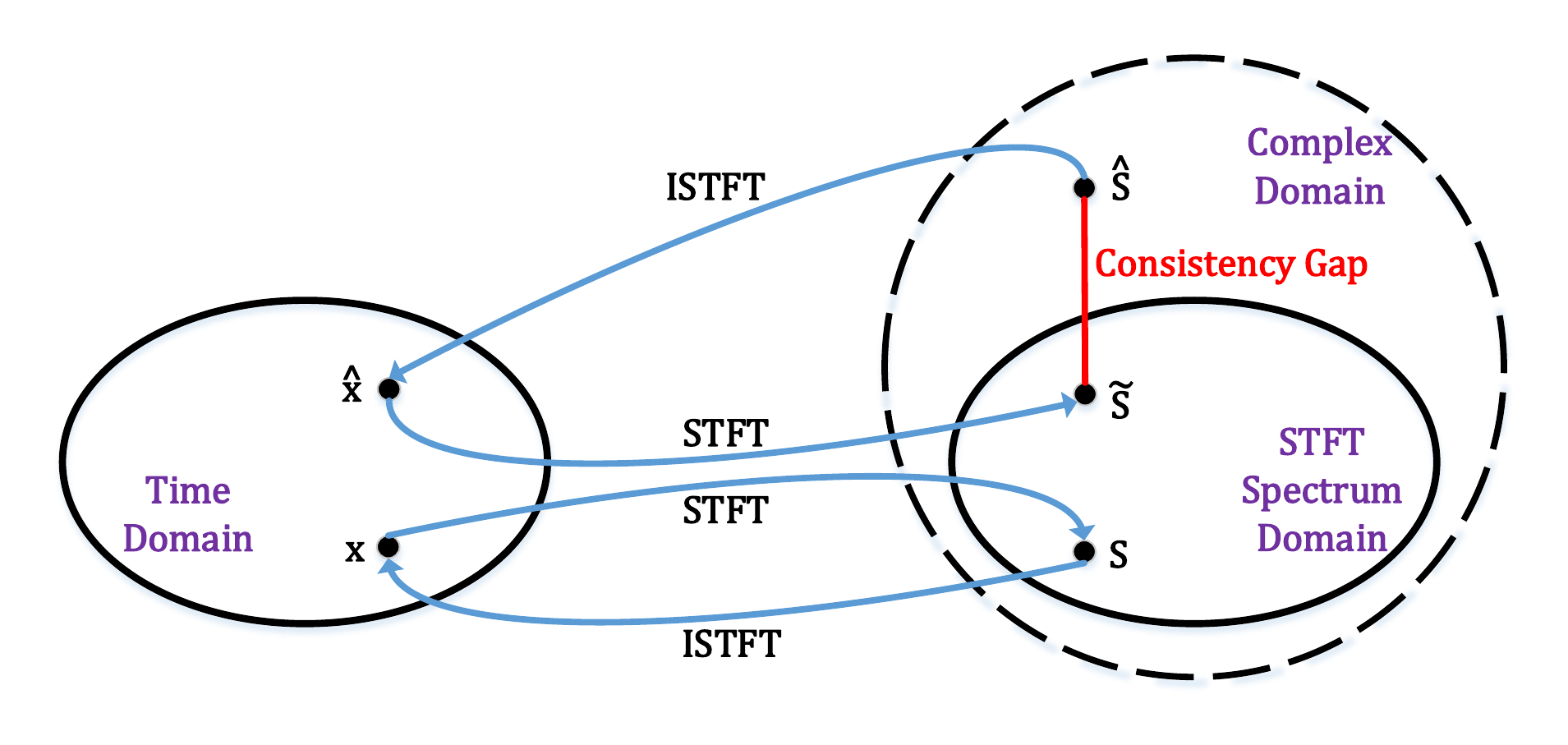}
    \caption{An interpretation diagram of the STFT spectrum consistency issue.
    }
    \label{fig: STFT_consistency}
\end{figure}

We first explain the short-time spectral inconsistency issue \cite{le2010fast}.
As shown in Figure \ref{fig: STFT_consistency}, the STFT spectrum domain $\mathbb{S}^{F\times \left(\frac{N}{2}+1\right)}$ is a subset of the complex domain $\mathbb{C}^{F\times \left(\frac{N}{2}+1\right)}$, i.e., $\mathbb{S}^{F\times \left(\frac{N}{2}+1\right)} \subset \mathbb{C}^{F\times \left(\frac{N}{2}+1\right)}$.
For a speech waveform $\bm{x}\in \mathbb{R}^T$, there must be
\begin{align}
\label{equ: consistency}
\bm{S}=STFT\left(\bm{x};L,\frac{T}{F},N\right)\in \mathbb{S}^{F\times \left(\frac{N}{2}+1\right)},\\
ISTFT\left(\bm{S};L,\frac{T}{F},N\right)=\bm{x}.
\end{align}
As shown in Figure \ref{fig: STFT_consistency}, assume that the reconstructed STFT spectrum $\hat{\bm{S}}\in \mathbb{C}^{F\times \left(\frac{N}{2}+1\right)}-\mathbb{S}^{F\times \left(\frac{N}{2}+1\right)}$ (i.e., Equation \ref{equ: whole process3}), there must be
\begin{align}
\label{equ: not consistency}
\hat{\bm{x}}=ISTFT\left(\hat{\bm{S}};L,\frac{T}{F},N\right),\\
STFT\left(\hat{\bm{x}};L,\frac{T}{F},N\right)=\tilde{\bm{S}}\in \mathbb{S}^{F\times \left(\frac{N}{2}+1\right)}\ne \hat{\bm{S}}.
\end{align}
At this point, an inconsistency problem occurs and the reconstructed $\hat{\bm{S}}$ is a ``fake" STFT spectrum.
The consistent spectrum corresponding to $\hat{\bm{S}}$ is $\tilde{\bm{S}}$.
The difference between them is the consistency gap, which needs to be narrowed.

To alleviate short-time spectral inconsistency error, we define the following loss functions.

\begin{itemize}
\item \textbf{STFT Consistency Loss}: The STFT consistency loss aims to narrow the gap between $\hat{\bm{S}}$ and its consistent spectrum $\tilde{\bm{S}}$ and it is calculated as follows:
    \begin{small}
    \begin{align}
    \label{equ: STFT Consistency Loss}
    \mathcal L_{C}=\mathbb{E}_{\left(\hat{\bm{S}},\tilde{\bm{S}}\right)} \overline{\left\{\left[Re(\hat{\bm{S}})-Re(\tilde{\bm{S}})\right]^2+\left[Im(\hat{\bm{S}})-Im(\tilde{\bm{S}})\right]^2\right\}}.
    \end{align}
    \end{small}

\item \textbf{Real and Imaginary Part Losses}: The real and imaginary part losses are proposed to narrow the gap between $\hat{\bm{S}}$ and its natural spectrum $\bm{S}$, and they are defined as the mean L1 distance of the real parts and imaginary parts, respectively, i.e.,
    \begin{align}
    \label{equ: Real and Imaginary Part Losses}
    \mathcal L_{R}=\mathbb{E}_{\left(\hat{\bm{S}},\bm{S} \right)} \overline{|Re(\hat{\bm{S}})-Re(\bm{S})|},\\
    \mathcal L_{I}=\mathbb{E}_{\left(\hat{\bm{S}},\bm{S} \right)} \overline{|Im(\hat{\bm{S}})-Im(\bm{S})|}.
    \end{align}

\end{itemize}

The total loss defined on reconstructed STFT spectra is
\begin{align}
\label{equ: STFT_loss_total}
\mathcal L_{S}=\mathcal L_{C}+\lambda_{RI}\mathcal L_{R}+\lambda_{RI}\mathcal L_{I},
\end{align}
where $\lambda_{RI}$ is hyperparameter.

\subsubsection{Losses defined on final waveforms}
\label{subsubsec3C4: Losses defined on final waveforms}

The losses defined on final waveforms aim to narrow the gap between the final synthesized waveform $\hat{\bm{x}}$ through ISTFT (i.e., Equation \ref{equ: whole process4}) and the natural one $\bm{x}$.
The losses are exactly the same as those used in HiFi-GAN \cite{kong2020hifi}, which adopts GAN-based training, including the GAN losses $\mathcal L_{GAN-G}$ and $\mathcal L_{GAN-D}$ for the generator and discriminator, respectively, feature matching loss $\mathcal L_{FM}$ and mel spectrogram loss $\mathcal L_{Mel}$.
The total loss defined on final waveforms is
\begin{align}
\label{equ: Final Loss W}
\mathcal L_{W}=L_{GAN-G}+\mathcal L_{FM}+\lambda_{Mel}\mathcal L_{Mel},
\end{align}
where $\lambda_{Mel}$ is hyperparameter.

\subsubsection{The final loss for model training}
\label{subsubsec3C5: The final loss for model training}

The training of the APNet vocoder follows the standard training process of GAN.
The final generator loss is a linear combination of multilevel losses, i.e.,
\begin{align}
\label{equ: Final Loss G}
\mathcal L_{G}=\lambda_{A}\mathcal L_{A}+\lambda_{P}\mathcal L_{P}+\lambda_{S}\mathcal L_{S}+\mathcal L_{W}.
\end{align}
where $\lambda_{A}$, $\lambda_{P}$ and $\lambda_{S}$ are hyperparameters.
The discriminator loss is $\mathcal L_{D}=\mathcal L_{GAN-D}$.

\subsubsection{The inference process}
\label{subsubsec3C6: The inference process}

The inference process strictly follows the all-frame-level calculation process from Equation \ref{equ: whole process1} to Equation \ref{equ: whole process4}, as shown in the black part of Figure \ref{fig: APNet} (i.e., ignore the gray part).

\section{Experiments}
\label{sec: Experiments}

\subsection{Experimental Setup}
\label{sec4A: Experimental Setup}

The LJSpeech dataset \cite{Ito2017LJ} was adopted in our experiments.
We chose 10480 and 1310 utterances to construct the training set and the validation set, respectively, and the remaining 1310 utterances were used as the test set.
The original waveforms in this dataset were downsampled to 16 kHz for our experiments.
When extracting features (e.g., mel spectrograms, amplitude spectra, phase spectra and F0) from natural waveforms for vocoders, the window size was 20 ms (i.e., $L=320$), the window shift was 5 ms (i.e., $\frac{T}{F}=80$), and the FFT point number was 1024 (i.e., $N=1024$).
For TTS experiments, we followed the open source implementation\footnote{\url{https://github.com/ming024/FastSpeech2}.} of a FastSpeech2-based acoustic model \cite{yasuda2019investigation}.

The subsequent experiments are organized as follows\footnote{Source codes are available at \url{https://github.com/yangai520/APNet}. Examples of generated speech can be found at \url{https://yangai520.github.io/APNet}.}.
We first compared our proposed APNet vocoder with other neural vocoders by analysis-synthesis (i.e., using natural acoustic features as input) and TTS (i.e., using acoustic features predicted by acoustic models as input) experiments.
Then, we further conducted ablation experiments on the APNet vocoder to explore the role of the proposed parallel phase estimation architecture and loss functions.

\subsection{Comparison among neural vocoders}
\label{sec4B: Comparison among neural vocoders}

\begin{table*}
\centering
    \caption{Objective and subjective evaluation results of \textbf{APNet}, \textbf{HiFi-GAN v1}, \textbf{HiFi-GAN v2}, \textbf{ISTFTNET}, \textbf{HiNet} and \textbf{MB-MelGAN} on the test sets of the LJSpeech dataset. Here, ``AS" and ``TTS" represent analysis-synthesis and TTS tasks respectively. ``$a\times$" represents $a\times$ real time. Objective results were calculated only on the analysis-synthesis task.}
    \resizebox{17.8cm}{1.3cm}{
    \begin{tabular}{c | c c c c c | c c | c c}
        \hline
        \hline
        & SNR(dB)$\uparrow$ & LAS-RMSE(dB)$\downarrow$ & MCD(dB)$\downarrow$ & F0-RMSE(cent)$\downarrow$ & V/UV error(\%)$\downarrow$ & RTF (CPU)$\downarrow$ & RTF (GPU)$\downarrow$ & MOS(AS)$\uparrow$ & MOS(TTS)$\uparrow$\\
        \hline
        \textbf{Natural Speech} & -- & -- & -- & -- & -- & -- & -- & 4.01$\pm$0.056 & 4.00$\pm$0.054 \\
        \hline
        \textbf{APNet} & 5.834 & 3.522 & 0.7729 & 19.83 & 3.142 & \textbf{0.068 (14.7$\times$)} & 0.0033 (303$\times$)& \textbf{3.93$\pm$0.057} & \textbf{3.90$\pm$0.053}\\
        \textbf{HiFi-GAN v1} & \textbf{7.528} & \textbf{3.207} & \textbf{0.5874} & \textbf{16.88} & \textbf{2.112} & 0.57 (1.76$\times$) & 0.0034 (291$\times$) & \textbf{3.99$\pm$0.056} & \textbf{3.90$\pm$0.060} \\
        \textbf{HiFi-GAN v2} & 4.937 & 4.721 & 1.150 & 36.71 & 4.164 &  0.30  (3.28$\times$)& 0.0027 (366$\times$) & 3.86$\pm$0.057 & 3.87$\pm$0.059 \\
        \textbf{ISTFTNET} & 4.824 & 4.905 & 1.252 & 38.04 & 4.490 & 0.077 (13.1$\times$) & \textbf{0.0012 (817$\times$)}& 3.87$\pm$0.060 & 3.86$\pm$0.056\\
        \textbf{HiNet} & 3.429 & 5.394 & 1.515 & 38.78 & 5.301 & 5.6 (0.178$\times$) & 0.22 (4.45$\times$) & 3.82$\pm$0.062 & -- \\
        \textbf{MB-MelGAN} & 3.669 & 6.744 & 2.077 & 40.11 & 7.029 & \textbf{0.054 (18.4$\times$)} & \textbf{0.0015 (672$\times$)} & 3.82$\pm$0.062 & 3.80$\pm$0.061\\

        \hline
        \hline
    \end{tabular}}
\label{tab_objective_subjective}
\end{table*}

We first conducted objective and subjective experiments to compare the performance of our proposed APNet vocoder and other neural vocoders.
The descriptions of these vocoders are as follows.
\begin{itemize}
\item {}{\textbf{APNet}}: The proposed APNet vocoder.
The input was 80-dimensional mel spectrograms.
The residual convolution network in both ASP and PSP consisted of 3 parallel blocks (i.e., $P=3$), and each block was formed by concatenating 3 subblocks (i.e., $Q=3$).
The kernel sizes of blocks were $k_1=3$, $k_2=7$ and $k_3=11$, and the dilation factors of subblocks within each block were $d_{*,1}=1$, $d_{*,2}=3$ and $d_{*,3}=5$.
The channel size of the convolution operations in both ASP and PSP was 512, except for output layers.
In particular, the channel size of the output layers was 513 (i.e., $\frac{N}{2}+1=513$).
The hyperparameters for loss functions were set as $\lambda_{A}=\lambda_{Mel}=45$, $\lambda_{P}=100$, $\lambda_{S}=20$ and $\lambda_{RI}=2.25$.
The model was trained using the AdamW optimizer \cite{loshchilov2018decoupled} with $\beta_1=0.8$ and $\beta_2=0.99$ on a single Nvidia 3090Ti GPU.
The learning rate decay was scheduled by a 0.999 factor in every epoch with an initial learning rate of 0.0002.
The batch size was 16, and the truncated waveform length was 8000 samples (i.e., 0.5 s) for each training step.
The model was trained until 3100 epochs.

\item {}{\textbf{HiFi-GAN v1}}: The v1 version of the HiFi-GAN vocoder \cite{kong2020hifi}.
The input was 80-dimensional mel spectrograms.
We reimplemented it using the open source implementation\footnote{\url{https://github.com/jik876/hifi-gan}.}.
We made small modifications to the open source code to fit our configurations.
The upsampling ratios were set as $h_1=5$, $h_2=4$, $h_3=2$ and $h_4=2$.
The training configuration is exactly the same as that of \textbf{APNet}.

\item {}{\textbf{HiFi-GAN v2}}: The v2 version of the HiFi-GAN vocoder \cite{kong2020hifi}.
Compared with \textbf{HiFi-GAN v1}, it just reduced the channel size of the convolution operations from 512 to 128.

\item {}{\textbf{ISTFTNET}}: The v2 version of the ISTFTNET vocoder \cite{kaneko2022istftnet}.
The input was 80-dimensional mel spectrograms.
We reimplemented it using the open source implementation\footnote{\url{https://github.com/rishikksh20/iSTFTNet-pytorch}.}.
We made small modifications to the open source code to fit our configurations, and it was designed based on \textbf{HiFi-GAN v2}.
The \textbf{ISTFTNET} passed 80-dimensional mel spectrograms through the first two residual networks incorporating upsampling layers of \textbf{HiFi-GAN v2} (i.e., $h_1=5$ and $h_2=4$) and then outputted the amplitude and phase spectra at a 4 kHz sampling rate.
The training configuration is exactly the same as that of \textbf{APNet}.

\item {}{\textbf{HiNet}}: The HiNet vocoder \cite{ai2020neural} we previously proposed.
The model configurations were the same as those used in the \textbf{Baseline-HiNet} model of our previous work \cite{ai2022denoising}.
The input included the 80-dimensional mel spectrograms, F0 extracted using YAPPT \cite{kasi2002yet}, and a voiced/unvoiced flag.

\item {}{\textbf{MB-MelGAN}}: The multi-band MelGAN vocoder \cite{yang2021multi}.
We reimplemented it using the open source implementation\footnote{\url{https://github.com/rishikksh20/melgan}.}.
We made small modifications to the open source code to fit our configurations.
The model generated 4 subband waveforms with a 4 kHz sampling rate, and the upsampling ratios for each band were set as 5, 2 and 2.
Finally, the full-band waveform was reconstructed by integrating these subband waveforms.
This method significantly improved the inference efficiency compared with the original MelGAN \cite{kumar2019melgan}.

\end{itemize}

We compared the performance of these vocoders using both objective and subjective evaluations.
Five objective metrics used in our previous work \cite{ai2020neural} were adopted here, including the signal-to-noise ratio (SNR), which reflected the distortion of waveforms, root MSE (RMSE) of log amplitude spectra (denoted by LAS-RMSE), which reflected the distortion in the frequency domain, mel-cepstrum distortion (MCD), which described the distortion of mel-cepstra, RMSE of F0 (denoted by F0-RMSE), which reflected the distortion of F0, and V/UV error, which was the ratio between the number of frames with mismatched V/UV flags and the total number of frames.
Among these metrics, SNR can be considered as an overall measurement of the distortions of both amplitude and phase spectra, while LAS-RMSE and MCD mainly present the distortion of amplitude spectra.
To evaluate the inference efficiency, the real-time factor (RTF), which is defined as the ratio between the time consumed to generate speech waveforms and the duration of the generated speech, was also utilized as an objective metric.
In our implementation, the RTF value was calculated as the ratio between the time consumed to generate all test sentences using a single Nvidia 3080Ti GPU or a single Intel Xeon E5-2680 CPU core and the total duration of the test set.

Regarding the subjective evaluation, mean opinion score (MOS) tests were conducted to compare the naturalness of these vocoders.
In each MOS test, twenty test utterances synthesized by vocoders along with the natural utterances were evaluated by at least 30 native English listeners on the crowdsourcing platform of Amazon Mechanical Turk\footnote{\url{https://www.mturk.com}.} with anti-cheating considerations \cite{buchholz2011crowdsourcing}.
Listeners were asked to give a naturalness score between 1 and 5, and the score interval was 0.5.
To compare the differences between the two vocoders individually, the ABX preference tests were also conducted on the Amazon Mechanical Turk platform and evaluated by at least 30 English native listeners.
In each ABX test, twenty utterances were randomly selected from the test set synthesised by two comparative vocoders.
The listeners were asked to judge which utterance in each pair had better speech quality or whether there was no preference.
In addition to calculating the average preference scores, the $p$-value of a $t$-test was used to measure the significance of the difference between two vocoders.

\begin{figure*}
    \centering
    \includegraphics[height=7cm]{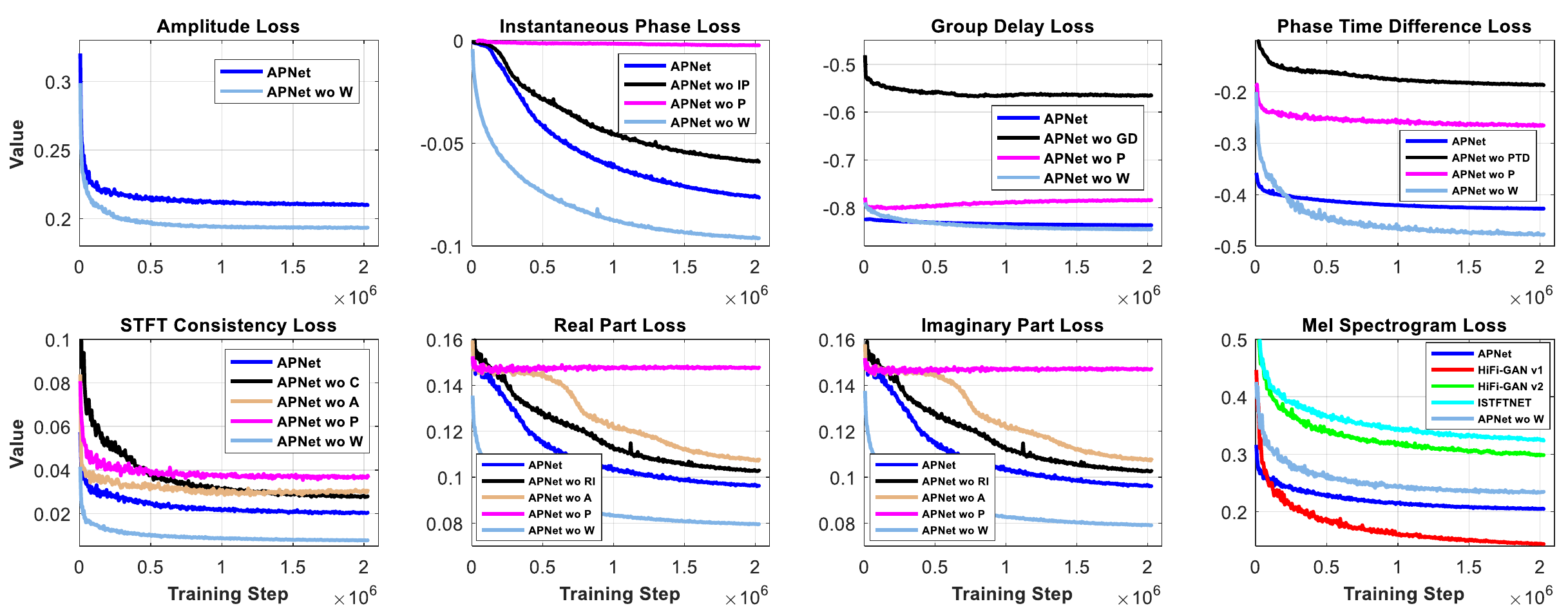}
    \caption{Loss curves for the validation set when training on the LJSpeech dataset.
    }
    \label{fig: Curve_LJSpeech}
\end{figure*}

The objective results on the test sets of the LJSpeech dataset for the analysis-synthesis task are listed in Table \ref{tab_objective_subjective}.
It is obvious that \textbf{HiFi-GAN v1} achieved the highest scores for all objective metrics.
Our proposed \textbf{APNet} outperformed \textbf{HiFi-GAN v2}, \textbf{ISTFTNET}, \textbf{HiNet} and \textbf{MB-MelGAN} on all objective metrics, which demonstrates the strong performance on synthesized speech quality of the proposed methods.
For more evidence, Figure \ref{fig: Curve_LJSpeech} draws the curves of the losses on the validation set as a function of training steps on the LJSpeech dataset.
The \textbf{APNet}, \textbf{HiFi-GAN v1}, \textbf{HiFi-GAN v2} and \textbf{ISTFTNET} are comparable because they all employ similar training modes and the common losses (e.g., the mel spectrogram loss) defined on the waveform.
Consistent with the previous observations, the mel spectrogram loss convergence value of \textbf{APNet} was higher than that of \textbf{HiFi-GAN v1} but significantly lower than that of \textbf{HiFi-GAN v2} and \textbf{ISTFTNET}.
Besides, we can see that all losses in \textbf{APNet} converged normally.


\begin{figure}
    \centering
    \includegraphics[height=6cm]{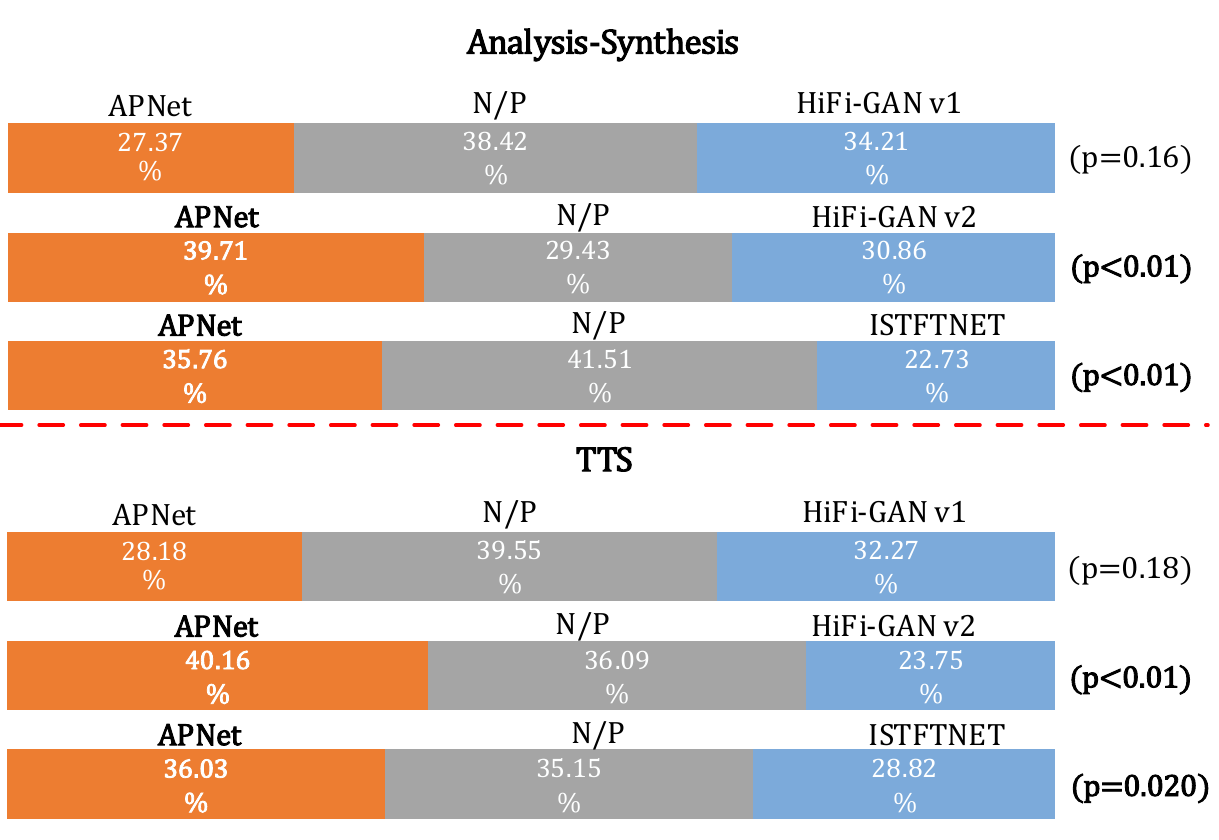}
    \caption{Average preference scores (\%) of ABX tests on speech quality between \textbf{APNet} and three other vocoders (i.e., \textbf{HiFi-GAN v1}, \textbf{HiFi-GAN v2} and \textbf{ISTFTNET}), where N/P stands for ``no preference" and $p$ denotes the $p$-value of a $t$-test between two vocoders.}
    \label{fig: ABX_main}
\end{figure}

The subjective MOS test results and RTFs on the test set of the LJSpeech dataset are also listed in Table \ref{tab_objective_subjective}.
Interestingly, the gap between our proposed \textbf{APNet} and the state-of-the-art \textbf{HiFi-GAN v1} was only 0.06 MOS scores for the analysis-synthesis task.
The MOS scores of \textbf{APNet} and \textbf{HiFi-GAN v1} were very close to those of ground truth.
For the TTS task, the \textbf{APNet} and \textbf{HiFi-GAN v1} had similar MOS scores, which indicated that the proposed APNet vocoder was robust to the acoustic features predicted by acoustic models.
Our proposed \textbf{APNet} also had better synthesized speech quality than \textbf{HiFi-GAN v2}, \textbf{ISTFTNET}, \textbf{HiNet} and \textbf{MB-MelGAN} regarding the MOS scores for both analysis-synthesis and TTS tasks.
The \textbf{HiNet} was discarded in the TTS task because it still required additional F0 input.
To further investigate the quality differences among \textbf{APNet}, \textbf{HiFi-GAN v1}, \textbf{HiFi-GAN v2} and \textbf{ISTFTNET}, we additionally performed several groups of ABX tests to compare the \textbf{APNet} with \textbf{HiFi-GAN v1}, \textbf{HiFi-GAN v2} and \textbf{ISTFTNET}, separately.
The results of the ABX tests are shown in Figure \ref{fig: ABX_main}.
The performance of \textbf{APNet} was not significantly different from that of \textbf{HiFi-GAN v1} ($p>0.05$) but significantly better than that of \textbf{HiFi-GAN v2} and \textbf{ISTFTNET} ($p<0.05$), both for analysis-synthesis and TTS tasks.
According to the listener's feedback, a few sentences of \textbf{HiFi-GAN v2} and \textbf{ISTFTNET} were slightly jittery or blurred.
Regarding the RTF, the \textbf{ISTFTNET} showed the fastest inference speed on GPU.
The inference speed of our proposed \textbf{APNet} on GPU was similar to the two versions of HiFi-GAN, and it achieved approximately 14x real-time inference on CPU, significantly faster than the two versions of HiFi-GAN.
The possible reason is that the CPU does not support parallel acceleration generation, resulting in slower direct waveform inference.
Although the \textbf{ISTFTNET} and \textbf{MB-MelGAN} had similar inference speeds on the CPU as \textbf{APNet}, their synthesized speech quality was noticeably worse.
The inference speed of \textbf{HiNet} was the slowest, either on GPU or CPU.
By comparing \textbf{HiNet} and \textbf{APNet}, we can see that the APNet vocoder has completely surpassed the HiNet vocoder in terms of both synthesized speech quality and inference efficiency, thanks to the introduction of the new frame-level PSP, as well as the joint training mode of ASP and PSP.
In conclusion, these experimental results confirm that our proposed APNet vocoder achieved synthesized speech quality comparable to HiFi-GAN but significantly improves the inference efficiency on CPU, benefiting from avoiding direct prediction of waveforms in our methods.

\begin{figure}
    \centering
    \includegraphics[height=7cm]{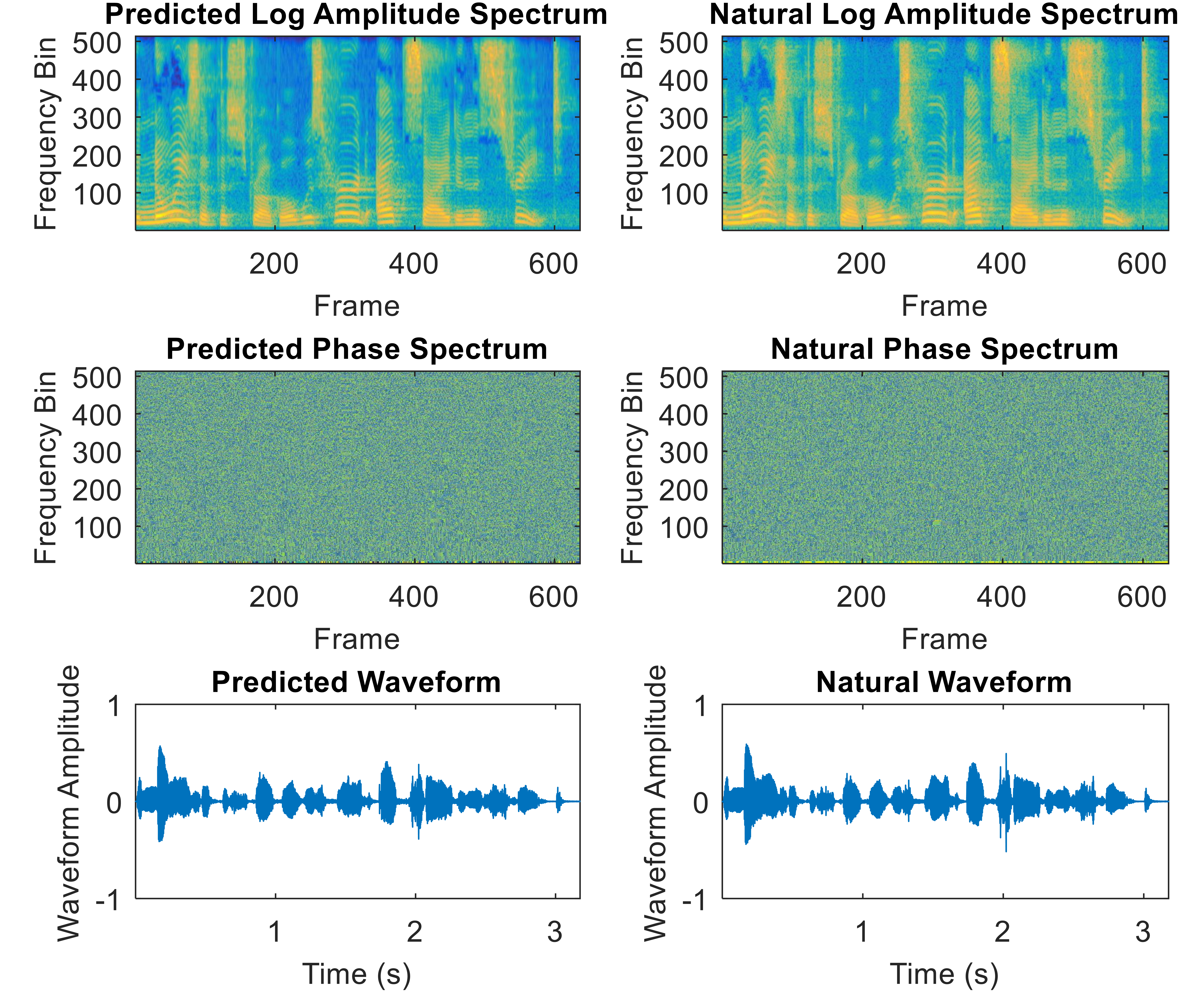}
    \caption{Visualized comparison of log amplitude spectrum, phase spectrum and waveform predicted by \textbf{APNet} with the natural ones for a test utterance.
    }
    \label{fig: Amplitude_phase_comparison}
\end{figure}

Figure \ref{fig: Amplitude_phase_comparison} draws the color maps of the log amplitude spectrum and wrapped phase spectrum predicted by the ASP and PSP in \textbf{APNet} and the reconstructed waveform for a test utterance.
The corresponding natural ones are also drawn in Figure \ref{fig: Amplitude_phase_comparison} for comparison.
Obviously, the predicted amplitude spectrum successfully recovered spectral information such as F0 and harmonics, but it was slightly over smooth compared to the natural amplitude spectrum.
Although it was difficult to see specific information from the color map of the phase spectrum, the overall distribution of the predicted phase spectrum was generally similar to the natural one by observing Figure \ref{fig: Amplitude_phase_comparison}.
In addition, the APNet vocoder also restored the overall waveform contour precisely compared with the natural one.

\subsection{Ablation studies}
\label{sec4C: Ablation studies}

\begin{table*}
\centering
    \caption{Objective evaluation results of \textbf{APNet} and its ablated variants on the test set of the LJSpeech dataset for the analysis-synthesis task.}
    \resizebox{18cm}{1.9cm}{
    \begin{tabular}{c | c | c c c c c}
        \hline
        \hline
         Descriptions & Vocoders & SNR(dB)$\uparrow$& LAS-RMSE(dB)$\downarrow$ & MCD(dB)$\downarrow$ & F0-RMSE(cent)$\downarrow$ & V/UV error(\%)$\downarrow$ \\
         \hline
          Original & \textbf{APNet} & 5.834 & 3.522 & 0.7729 & 19.83 & 3.142\\
         \hline
         Ablate Parallel Phase Estimation Architecture & \textbf{APNet wo PPEA} & 3.811 & \textbf{3.199} & 0.9967 & 38.35 & 5.842\\
          \hline
         Ablate Amplitude-related Loss & \textbf{APNet wo A} & 5.289 & 3.708 & 0.9382 & 25.41 & 3.696\\
         \hline
         \multirow{4}{*}{Ablate Phase-related Losses}
         & \textbf{APNet wo IP} & 5.605 & 3.699 & 0.8188 & 20.83 & 3.587\\
         & \textbf{APNet wo GD} & 6.323 & 4.091 & 0.9630 & 18.93 & 3.127\\
         & \textbf{APNet wo PTD} & 5.295 & 3.434 & 0.8843 & 26.71 & 4.606\\
         & \textbf{APNet wo P} & 4.163 & 3.683 & 0.9920 & 40.51 & 5.821\\
         \cline{1-7}
         \hline
         \multirow{2}{*}{Ablate STFT Spectrum-related Losses}
         & \textbf{APNet wo C} & 6.077 & 3.712 & 0.7884 & 19.09 & 2.916\\
         & \textbf{APNet wo RI} & 5.849 & 3.681 & \textbf{0.7668} & 20.58 & 3.366\\
         \cline{1-7}
         \hline
         Ablate Waveform-related Losses & \textbf{APNet wo W} & \textbf{6.447} & 5.421 & 1.269 & \textbf{14.81} & \textbf{2.390}\\
         \hline
        \hline
    \end{tabular}}
\label{tab_objective_ablation}
\end{table*}

We then conducted several ablation experiments to explore the roles of the proposed parallel phase estimation architecture and loss functions in the APNet vocoder.
We utilized some of the objective metrics used in the previous section and the subjective ABX preference test to compare \textbf{APNet} with its ablated variants.
The objective and subjective results are shown in Table \ref{tab_objective_ablation} and Figure \ref{fig: ABX}, respectively.
Ablation experiments were performed only on the analysis-synthesis task.

\subsubsection{Ablate the parallel phase estimation architecture}
\label{subsec4C1: Ablate parallel phase estimation architecture}

Here is one ablated vocoder for comparison with \textbf{APNet}, i.e.,

\begin{itemize}
\item {}{\textbf{APNet wo PPEA}}: Removing the parallel phase estimation architecture (i.e., two parallel output layers and phase calculation formula $\bm{\Phi}$) from \textbf{APNet}.
    The output of the residual convolution network in the PSP passes through a linear layer without activation to predict the phase spectra.
\end{itemize}

\begin{figure}
    \centering
    \includegraphics[height=7cm]{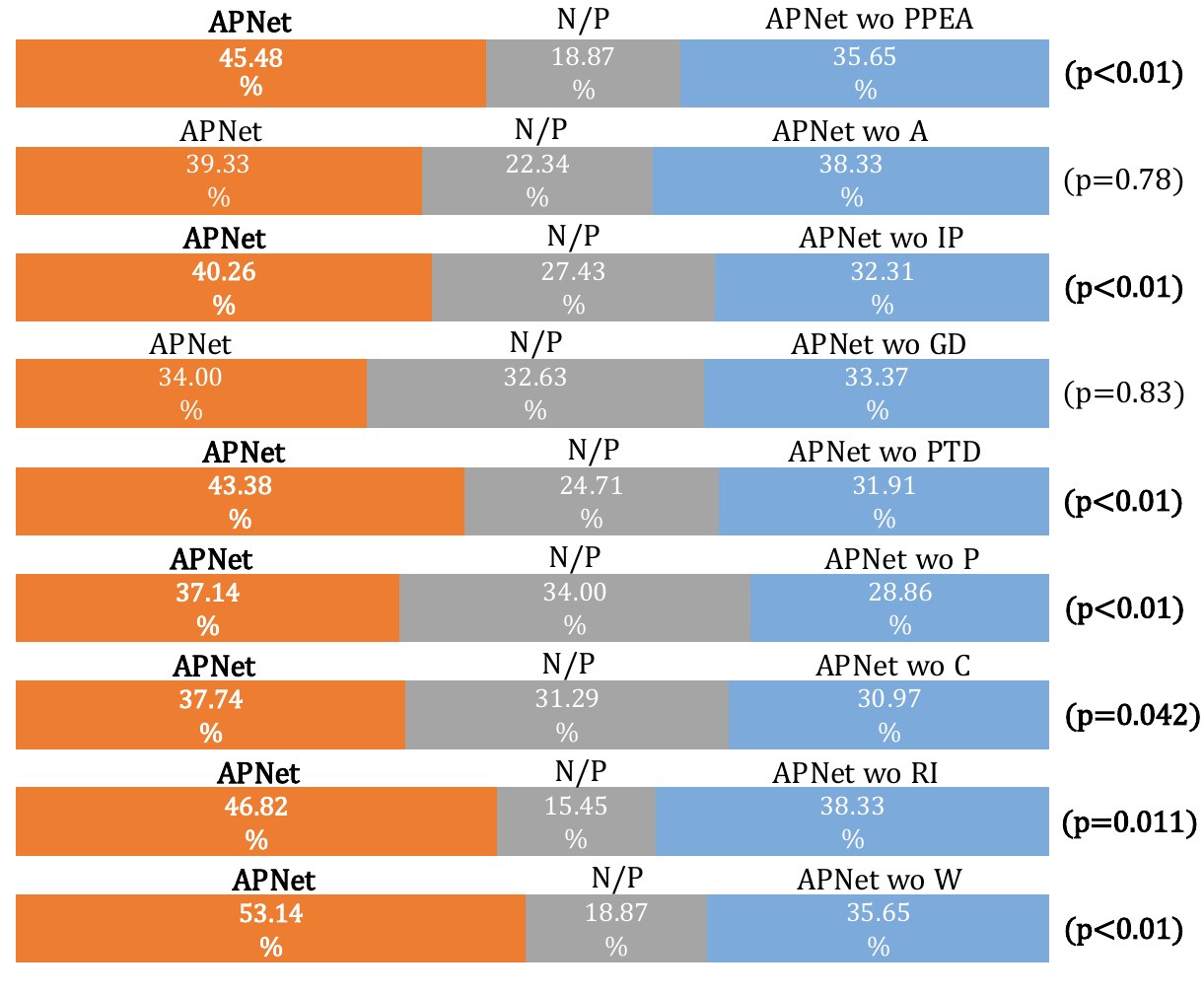}
    \caption{Average preference scores (\%) of ABX tests on speech quality between \textbf{APNet} and its ablated variants, where N/P stands for ``no preference" and $p$ denotes the $p$-value of a $t$-test between two vocoders.}
    \label{fig: ABX}
\end{figure}

\begin{figure}
    \centering
    \includegraphics[height=4.6cm]{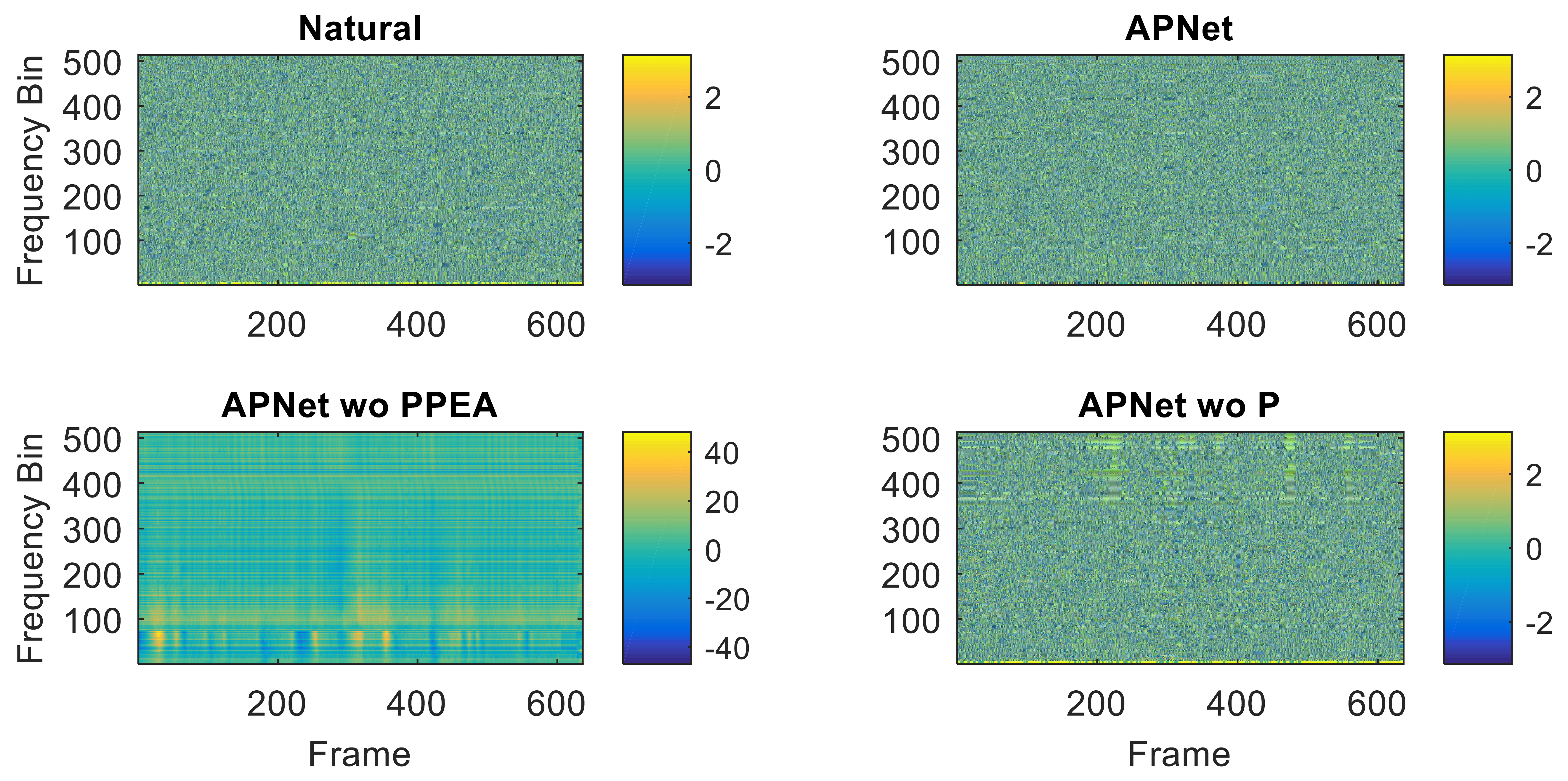}
    \caption{A comparison among the natural phase spectrum and phase spectra predicted by the PSP in \textbf{APNet}, \textbf{APNet wo PPEA} and \textbf{APNet wo P} for a test utterance.}
    \label{fig: Phase_comparsion}
\end{figure}

Unsurprisingly, the \textbf{APNet} outperformed \textbf{APNet wo PPEA} very significantly according to both objective and subjective results ($p<0.01$) in Table \ref{tab_objective_ablation} and Figure \ref{fig: ABX}.
In Figure \ref{fig: Phase_comparsion}, we drew the color maps of the phase spectra predicted by the PSP in \textbf{APNet} and \textbf{APNet wo PPEA}.
Obviously, the \textbf{APNet wo PPEA} failed to control the range of the predicted phase which overflowed the principal value interval $(-\pi,\pi]$.
This led to a loss of phase prediction accuracy and degradation of synthesized speech quality, thus confirming the effectiveness of our proposed parallel phase estimation architecture.

\subsubsection{Ablate the amplitude-related loss}
\label{subsec4C2: Ablate amplitude-related loss}

Here is one ablated vocoder for comparison with \textbf{APNet}, i.e.,

\begin{itemize}
\item {}{\textbf{APNet wo A}}: Removing the amplitude loss $\mathcal L_{A}$ from \textbf{APNet} at the training stage.
\end{itemize}

\begin{figure}
    \centering
    \includegraphics[height=8.5cm]{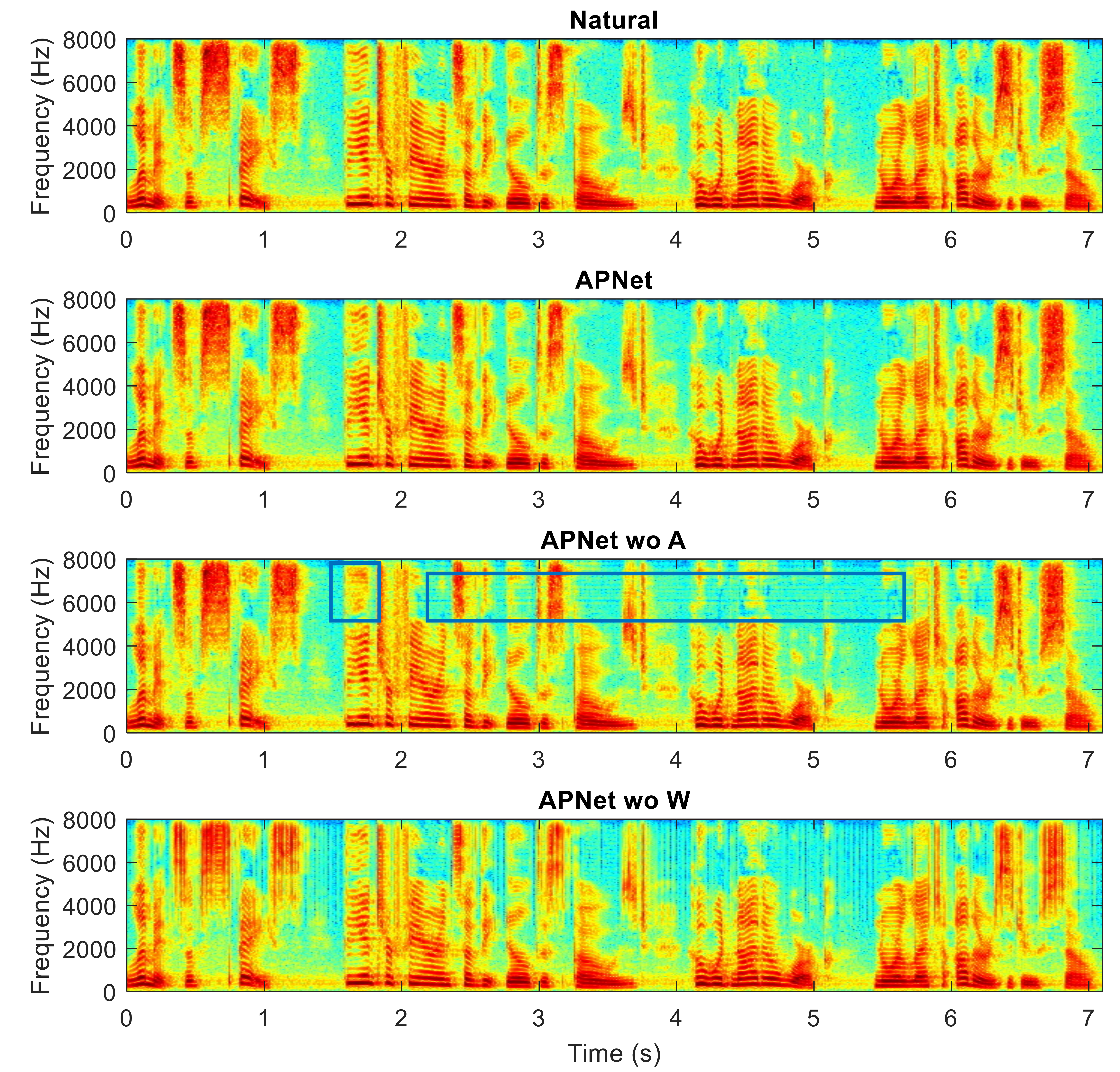}
    \caption{A comparison among the spectrograms of the natural speech and speeches generated by \textbf{APNet}, \textbf{APNet wo A} and \textbf{APNet wo W}.}
    \label{fig: Spectrogram_issue}
\end{figure}

Table \ref{tab_objective_ablation} shows that the \textbf{APNet wo A} attenuated all objective metrics compared with \textbf{APNet}.
In addition, the amplitude loss $\mathcal L_{A}$ of \textbf{APNet wo A} on the validation set did not converge and stabilized at approximately 8.5, while that of \textbf{APNet} converged to approximately 0.2, as shown in Figure \ref{fig: Curve_LJSpeech}.
From Figure \ref{fig: Curve_LJSpeech}, we can also see that removing $\mathcal L_{A}$ also negatively affected the convergence of the STFT spectrum-related losses (i.e., $\mathcal L_{C}$, $\mathcal L_{R}$ and $\mathcal L_{I}$).
However, there was no significant subjective difference ($p>0.05$) between \textbf{APNet} and \textbf{APNet wo A} considering the sense of hearing, as shown in Figure \ref{fig: ABX}.
For more evidence, Figure \ref{fig: Spectrogram_issue} draws the spectrograms of the speech generated by \textbf{APNet} and \textbf{APNet wo A}.
Obviously, the spectrogram of \textbf{APNet wo A} exhibited high-frequency horizontal streaks (see the blue boxes in Figure \ref{fig: Spectrogram_issue}), which may degrade the objective results.
However, this issue was not sensitive to hearing according to the subjective results.

\subsubsection{Ablate the phase-related losses}
\label{subsec4C3: Ablate phase-related losses}

Here are four ablated vocoders for comparison with \textbf{APNet}, i.e.,

\begin{itemize}
\item {}{\textbf{APNet wo IP}}: Removing the instantaneous phase loss $\mathcal L_{IP}$ from \textbf{APNet} at the training stage.
\item {}{\textbf{APNet wo GD}}: Removing the group delay loss $\mathcal L_{GD}$ from \textbf{APNet} at the training stage.
\item {}{\textbf{APNet wo PTD}}: Removing the phase time difference loss $\mathcal L_{PTD}$ from \textbf{APNet} at the training stage.
\item {}{\textbf{APNet wo P}}: Removing $\mathcal L_{IP}$, $\mathcal L_{GD}$ and $\mathcal L_{PTD}$ from \textbf{APNet} at the training stage (i.e., no losses directly defined on phase spectra).
\end{itemize}

From Table \ref{tab_objective_ablation}, we can see that the \textbf{APNet wo IP} and \textbf{APNet wo PTD} degraded most objective metrics.
However, the \textbf{APNet wo GD} even improved the SNR and F0-related metrics, which requires further investigation.
Removing all the phase-related losses (i.e., \textbf{APNet wo P}) caused a sharp degradation in all objective metrics.
Considering the convergence curves shown in Figure \ref{fig: Curve_LJSpeech}, we can see that removing a loss function at the training stage was bound to negatively affect the convergence of this loss value on the validation set.
The instantaneous phase loss curves of the \textbf{APNet wo IP} and \textbf{APNet wo P} also suggested that the phase difference losses (i.e., $\mathcal L_{GD}$ and $\mathcal L_{PTD}$) had a positive effect on the instantaneous phase loss (i.e., $\mathcal L_{IP}$).
Interestingly, the two phase difference losses were mutually exclusive by comparing the \textbf{APNet wo P} with \textbf{APNet wo GD} and \textbf{APNet wo PTD} using the group delay loss curve and phase time difference loss curve, respectively in Figure \ref{fig: Curve_LJSpeech}.
In addition, removing phase-related losses also significantly affected the convergence of the STFT spectrum-related losses (i.e., $\mathcal L_{C}$, $\mathcal L_{R}$ and $\mathcal L_{I}$).
It is reasonable that the amplitude- and phase-related losses facilitated the convergence of the STFT spectrum-related losses.
By comparing the \textbf{APNet wo A} and \textbf{APNet wo P}, we can also confirm that the phase-related losses contributed more to improving STFT spectrum consistency than the amplitude-related loss.

\begin{figure}
    \centering
    \includegraphics[height=6.5cm]{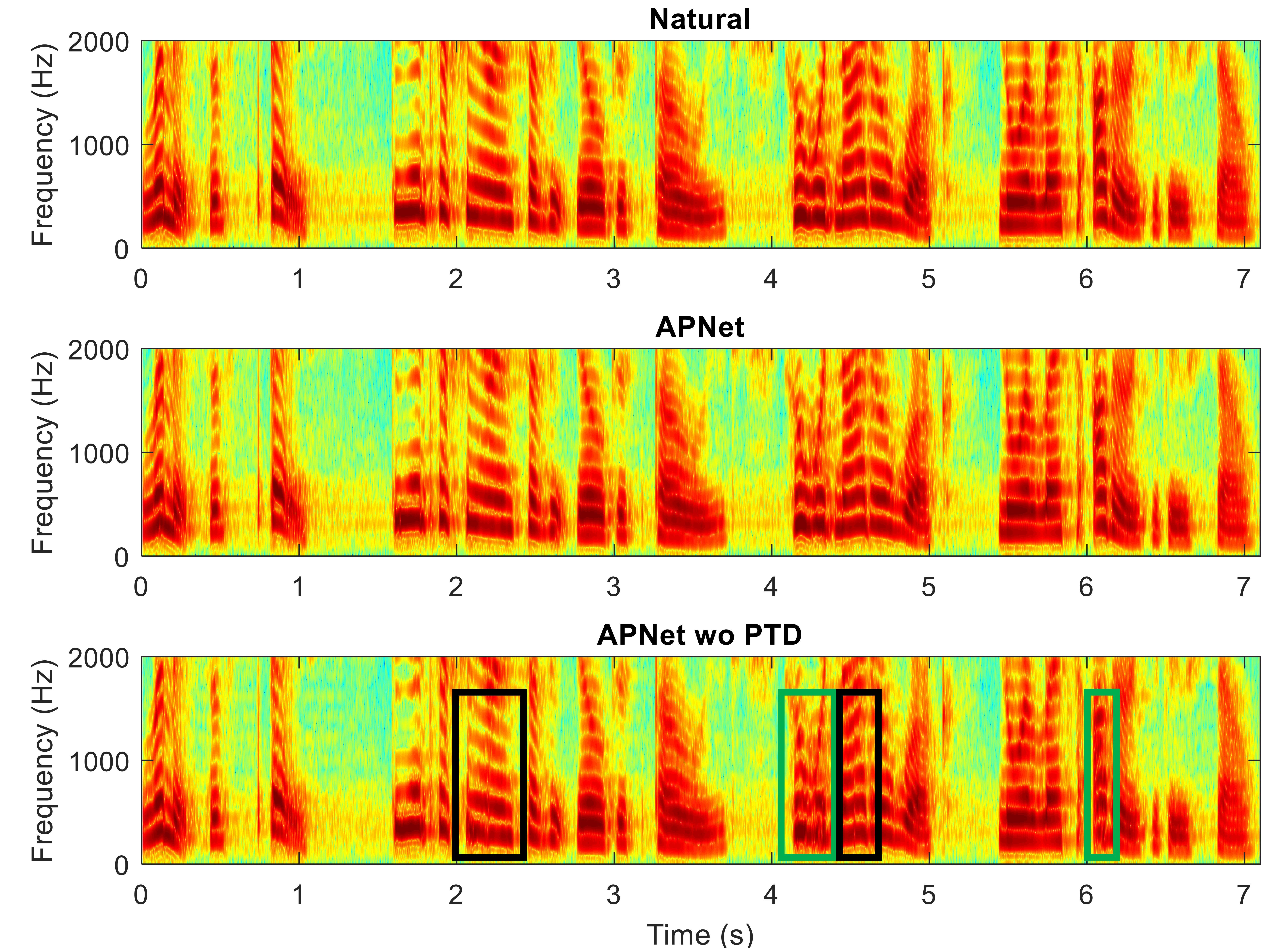}
    \caption{A comparison between the spectrograms of the natural speech and speeches generated by \textbf{APNet} and \textbf{APNet wo PTD} for the interpretation of the spectral discontinuity.}
    \label{fig: Discontinuity}
\end{figure}

Regarding the subjective results in Figure \ref{fig: ABX}, the \textbf{APNet wo IP}, \textbf{APNet wo PTD} and \textbf{APNet wo P} outperformed the \textbf{APNet} very significantly ($p<0.01$).
However, there was no significant difference ($p>0.05$) between the \textbf{APNet} and \textbf{APNet wo GD}.
Although the \textbf{APNet wo GD} significantly degraded the amplitude spectra according to the LAS-RMSE and MCD values in Table \ref{tab_objective_ablation}, the amplitude distortion might not cause a significant decrease in the sense of hearing.
This finding was similar to the conclusion in subsection \ref{subsec4C2: Ablate amplitude-related loss}.
These results suggested that the degree of importance of the three losses was $\mathcal L_{PTD}>\mathcal L_{IP}>\mathcal L_{GD}$.
We can also draw the conclusion that these three losses should be combined and balanced during training, which was more helpful to improve the phase prediction ability of PSP, rather than using any of these losses alone.
In addition, we found that a common issue existing in these four ablated vocoders was \emph{spectral discontinuity}.
We took the \textbf{APNet wo PTD} as an example and drew the spectrograms in Figure \ref{fig: Discontinuity}.
Discontinuity issues usually occurred at low frequencies, mainly including spectral blurring (green boxes in Figure \ref{fig: Discontinuity}) and spectral splitting (black boxes in Figure \ref{fig: Discontinuity}), resulting in slurred and clicky pronunciation.
This was due to the inconsistency of the STFT spectrum reconstructed by the predicted amplitude and phase spectra.
Therefore, phase-related losses facilitated the alleviation of STFT spectral inconsistency.
Finally, we also compared the phase spectra predicted by the PSP in \textbf{APNet} and \textbf{APNet wo P} in Figure \ref{fig: Phase_comparsion}.
Obviously, there existed horizontal stripes in the phase spectra of \textbf{APNet wo P} at the high frequency band, resulting in the degradation of phase prediction accuracy, which further demonstrated the effectiveness of our proposed phase-related losses.

\subsubsection{Ablate the STFT spectrum-related losses}
\label{subsec4C4: Ablate STFT spectrum-related losses}

Here are two ablated vocoders for comparison with \textbf{APNet}, i.e.,

\begin{itemize}
\item {}{\textbf{APNet wo C}}: Removing the STFT consistency loss $\mathcal L_{C}$ from \textbf{APNet} at the training stage.
\item {}{\textbf{APNet wo RI}}: Removing the real part loss $\mathcal L_{R}$ and imaginary part loss $\mathcal L_{I}$ from \textbf{APNet} at the training stage.
\end{itemize}

From Table \ref{tab_objective_ablation}, we can see that the objective results of the \textbf{APNet wo C} and \textbf{APNet wo RI} are similar to those of the \textbf{APNet}, which requires further investigation.
By observing Figure \ref{fig: Curve_LJSpeech}, although removing losses $\mathcal L_{C}$, $\mathcal L_{R}$ and $\mathcal L_{I}$ during the training stage caused an increase in the corresponding losses on the validation set, their increase degrees were smaller than that of the phase-related losses (i.e., $\mathcal L_{IP}$, $\mathcal L_{GD}$ and $\mathcal L_{GDT}$).
This suggested that the STFT spectrum-related losses played an auxiliary role in improving the consistency of the STFT spectra.
Regarding the subjective results shown in Figure \ref{fig: ABX}, the \textbf{APNet} outperformed both the \textbf{APNet wo C} and \textbf{APNet wo RI} significantly ($p<0.05$).
Other findings were that the \textbf{APNet wo C} and \textbf{APNet wo RI} also exhibited spectral discontinuity issues, as shown in Figure \ref{fig: Discontinuity}, which may be the reason for the decreased sense of hearing.
Therefore, the STFT spectrum-related losses were effective for improving the synthesized speech quality and alleviating the STFT spectral inconsistency issues.

\subsubsection{Ablate the waveform-related losses}
\label{subsec4C5: Ablate waveform-related losses}

Here is one ablated vocoder for comparison with \textbf{APNet}, i.e.,

\begin{itemize}
\item {}{\textbf{APNet wo W}}: Removing all losses defined on the final waveform (i.e., $L_{GAN-G}$, $L_{GAN-D}$, $L_{FM}$ and $L_{Mel}$) from \textbf{APNet} at the training stage.
\end{itemize}

Interestingly, the \textbf{APNet wo W} degraded the LAS-RMSE and MCD very significantly, but it improved the other three objective metrics, which requires further investigation.
By observing Figure \ref{fig: Curve_LJSpeech}, we found that the \textbf{APNet wo W} facilitated the convergence of all losses except the mel spectrogram loss.
This finding means that waveform-related losses restricted the independent learning of the ASP and PSP in \textbf{APNet}.
Removing waveform-related losses pushed the ASP and PSP to local optima but ignored the global optima of the final reconstructed waveforms.
As expected, the \textbf{APNet} outperformed \textbf{APNet wo W} very significantly ($p<0.01$) according to the subjective results in Figure \ref{fig: ABX}.
Figure \ref{fig: Spectrogram_issue} also draws the spectrograms of the speech generated by \textbf{APNet wo W}.
There were obvious vertical lines existing in the spectrogram of \textbf{APNet wo W}, which caused annoying clicking noises and degraded the amplitude-related objective results (i.e., the LAS-RMSE and MCD) and subjective results.
However, the vertical line issue did not destroy F0, as shown in Figure \ref{fig: Spectrogram_issue}, which may be the reason why the objective metrics related to F0 have not deteriorated.
However, after removing waveform-related losses, the model training efficiency was improved by approximately 8 times.
This finding is encouraging and points the way for our future work to improve the training efficiency of the APNet vocoder.

\section{Conclusion}
\label{sec: Conclusion}

In this paper, we proposed a novel neural vocoder called APNet, which incorporates the direct prediction of speech amplitude and phase spectra.
The APNet vocoder consists of an ASP and a PSP, which both perform at the frame level.
The ASP employs a residual convolution network to predict the log amplitude spectra from input acoustic features.
The PSP introduces a parallel phase estimation architecture on the basis of ASP and successfully realizes direct phase spectrum prediction from acoustic features.
Some well-designed loss functions are defined on the predicted log amplitude spectra, phase spectra, reconstructed STFT spectra and final waveforms, and jointly train the ASP and PSP.
Experimental results show that our proposed APNet vocoder achieved approximately  14x real-time inference efficiency on a CPU, which was 8x faster than HiFi-GAN v1 when its synthesized speech quality was comparable to HiFi-GAN v1.
Ablation studies also demonstrated the effectiveness of the proposed parallel phase estimation architecture and loss functions.

In future work, we will 1) adapt the APNet vocoder to other configurations (e.g., increase the frame shift, increase the waveform sampling rate, etc.); 2) improve the training efficiency of the APNet vocoder; and 3) apply the proposed phase prediction method to other speech generation tasks, such as speech enhancement (SE), speech bandwidth extension (BWE), etc.

\bibliographystyle{IEEEtran}
\bibliography{mybib}
\end{document}